\documentstyle[12pt,fleqn]{article}
\begin{document}
\baselineskip=\normalbaselineskip
\multiply\baselineskip by 150 \divide\baselineskip by 100
\parskip10pt
\pagenumbering{arabic}
\begin{titlepage}
\begin{flushright}
{MSUHEP-50501\\May, 1995}
\end{flushright}
\vspace{6mm}
\begin{center}
{\bf Studying Anomalous $WW\gamma$ and $WWZ$ Couplings}\\
{\bf with Polarized $p\bar{p}$ Collisions}\\
\vspace{6mm}
\end{center}
\begin{center}
{Michael Wiest, Daniel R. Stump, Douglas O. Carlson, C.--P. Yuan}\\
{Department of Physics and Astronomy}\\
{Michigan State University}\\
{East Lansing, MI 48824}\\
\end{center}
\vspace{6mm}
\raggedbottom
\setcounter{page}{1}
\relax
\begin{abstract}
\noindent
We calculate tree-level cross-sections for $W^{+}\gamma$ and $W^{+}W^{-}$
production in proton-antiproton collisions, with one $W$ decaying to
leptons, with  anomalous electroweak triple-boson
coupling parameters $\Delta\kappa$ and $\lambda$.
We compare the unpolarized cross-sections to those for a polarized
proton beam, to study how a polarized proton beam would improve
experimental tests of anomalous couplings.
\end{abstract}
\vspace{2cm}
\end{titlepage}
\newpage
\section{Introduction}

In the last few years the SPIN collaboration has shown
in various technical notes that it is feasible to polarize
the proton beam, longitudinally or transversely, during the
colliding mode of the Tevatron \cite{SPIN}.
Taking proton polarization as a possibility, we examine
one of the possible physics topics that could be pursued
with such a beam configuration -- to study tri-boson couplings
of the weak gauge bosons.
(Other interesting physics topics involving polarization
at the Tevatron collider can be found in Ref.~\cite{eoipol}.)

An important test of the electroweak Standard Model is to measure
the couplings among gauge fields.  In this paper we are concerned
with the $WW\gamma$ and $WWZ$ couplings.  Let $W^{\mu}(x)$, $Z^{\mu}(x)$,
and $A^{\mu}(x)$ denote the fields of $W^{-}$, $Z^{0}$, and $\gamma$;
then the interaction Lagrangian we consider is
\begin{equation}
{\cal{L}}_{3}=
-ig(W^{\dagger}_{\mu\nu}W^{\mu}-W_{\mu\nu}W^{\dagger\mu})
(A^{\nu}\sin\theta_{W}+Z^{\nu}\cos\theta_{W})
\end{equation}
\[-igW^{\dagger}_{\mu}W_{\nu}
(A^{\mu\nu}\kappa_{\gamma}\sin\theta_{W}
+Z^{\mu\nu}\kappa_{Z}\cos\theta_{W})\]
\[-\frac{ig}{M_{W}^{2}}W^{\dagger}_{\mu\alpha}W^{\alpha}_{\nu}
(A^{\mu\nu}\lambda_{\gamma}\sin\theta_{W}\
+Z^{\mu\nu}\lambda_{Z}\cos\theta_{W})\]\noindent
where $A_{\mu\nu}=\partial_{\mu}A_{\nu}-\partial_{\nu}A_{\mu}$, {\it etc.}
${\cal{L}}_{3}$ is gauge invariant with respect to
U(1) electromagnetic gauge transformations.
The parameter $\kappa_{\gamma}$ is the anomalous magnetic moment
of the $W^{-}$, as defined by Lee and Yang \cite{LY};
$\lambda_{\gamma}$ is an anomalous electric quadrupole moment.
The parameters $\kappa_{Z}$
and $\lambda_{Z}$ are similar $WWZ$ couplings.
In the Standard SU(2)$\times$U(1) gauge theory these coupling
parameters have the definite values
\begin{equation}
\kappa_{\gamma}=\kappa_{Z}=1~,~~\lambda_{\gamma}=\lambda_{Z}=0~~
{\rm (Standard~Model).}
\end{equation}
\noindent
We define $\Delta\kappa$ by
\begin{equation}
\Delta\kappa=\kappa-1~.\
\end{equation}\noindent
If $\Delta\kappa$ or $\lambda$ is significantly
different than 0 for either
$\gamma$ or $Z^{0}$, then the Standard  SU(2)$\times$U(1) gauge
theory is not the complete theory of the electroweak interactions.
In the Standard Model, $\Delta\kappa$ and $\lambda$ can be
induced at loop level, but only of the size
of ${\cal O}(g^{2}/16\pi^{2})$ at one-loop level
from naive dimensional analysis \cite{Georgi}.
Thus setting experimental limits on the anomalous $WW\gamma$ or
$WWZ$ couplings is an important test of the Standard Model;
actually discovering large anomalous interactions would be
a sign of new physics.
We treat ${\cal{L}}_{3}$ as an effective Lagrangian and only use it for
tree-level calculations.

To determine the $WWV$ couplings (where $V$ stands for
$\gamma$ or $Z^{0}$) it is necessary to measure the experimental
cross-section, or the distribution of some kinematic variable,
for a process that depends on the $WWV$ coupling,  and compare the
measurement to a calculated prediction.
In this paper we consider two processes in
proton-antiproton collisions
\[p+\bar{p} \rightarrow W^{+}\left(\rightarrow \ell^{+}~\nu_{\ell}\right)
+\gamma~,\]
\[p+\bar{p} \rightarrow W^{\pm}
\left(\rightarrow \ell^{\pm}~{\nu\!\!\!\!^{^{(-)}}}\!\!_{\ell}\right)
+W^{\mp}\left(\rightarrow 2 {\it ~jets}\right)~.\]\noindent
For $\ell$ we include both $e$ and $\mu$.
The unpolarized cross-section for $p\bar{p}\rightarrow W^{-}\gamma$
is equal to that for  $p\bar{p}\rightarrow W^{+}\gamma$ in a CP
invariant theory such as ${\cal L}_{3}$; however, we are
interested in polarized scattering, for which $W^{+}\gamma$ is
more interesting than $W^{-}\gamma$ for a $p\bar{p}$
collider with a polarized proton beam.

The purpose of this paper is to explore the experimental
search for anomalous couplings in proton-antiproton collisions,
{\em assuming the protons are longitudinally polarized.}
The antiprotons are assumed to be unpolarized.
With the Tevatron collider in mind \cite{SPIN,eoipol},
we consider the center-of-mass energy equal to 2 TeV.

The reaction cross-section for a process involving the $WW\gamma$ or
$WWZ$ coupling depends on the longitudinal polarization of the proton
through spin-dependent parton distribution functions.
The coupling of $W^{\pm}$ to quarks ($ud$ or other flavor combinations)
is a V$-$A interaction, so the parton-level cross-section depends
strongly on the helicities of the quarks: for massless quarks
a $W^{\pm}$ couples only to left-handed ($L$) quarks and right-handed
($R$) antiquarks.
Thus the parton-level process depends strongly on helicity.
The question is whether the {\em proton process} depends
strongly on proton helicity.
If a polarized proton contained equal parton densities
of left-handed and right-handed quarks, then the proton
cross-section would
not depend on the proton helicity.  However, we know that
the densities of $L$ and $R$ quarks are not equal for polarized
protons.  Therefore, the
$p\bar{p}$ cross-section will be different for left-polarized and
right-polarized protons.

Our calculations of the polarized-proton cross-sections depend on
polarized parton distribution functions (hereafter abbreviated
ppdf's), and these are only known with limited accuracy.
The ppdf's are defined as follows: For any parton type $f$, we define
\begin{equation}
f_{+}(x)=\frac{1}{2}(f(x)+\Delta{f}(x))
\end{equation}
\[~~~~~~~~~~={\rm density~of~L~(or~R)~parton~in~L~(or~R)~proton}~,\]
\noindent
\begin{equation}
f_{-}(x)=\frac{1}{2}(f(x)-\Delta{f}(x))
\end{equation}
\[~~~~~~~~~~={\rm density~of~L~(or~R)~parton~in~R~(or~L)~proton}~,\]
\noindent
where $x$ is the momentum fraction of the parton.  There are nine
different parton types
\begin{equation}
f=u_{val}~,~d_{val}~,~u_{sea}=\bar{u}_{sea}~,~d_{sea}=\bar{d}_{sea}~,
{}~g~,~s=\bar{s}~,~c=\bar{c}~,~b=\bar{b}~,~t=\bar{t}~.
\end{equation}\noindent
In the ppdf's we used, the $u$ and $d$ sea distributions are equal,
but different than the $s$ and $c$ distributions, and the $b$ and $t$
distributions are zero
\begin{equation}
u_{sea}(x)=d_{sea}(x)~;~b(x)=0~;~t(x)=0~.\noindent
\end{equation}

Figure 1.1 shows the polarization dependence of the ppdf's we used,
by plotting $x\,\Delta{f}(x)$ for several parton species.  The ppdf's
depend on momentum scale $Q$; {\it i.e.} $f_{\pm}=f_{\pm}(x,Q^{2})$.
Fig.\ 1.1 corresponds to $Q$=80 GeV.
(These ppdf's are calculated from a program based on
Morfin-Tung parton distribution functions \cite{MT,Ladinsky}.)
The ppdf's have been measured, to some limited precision,
from polarized deep-inelastic lepton scattering \cite{polpdf}.
Recent data from the Spin Muon Collaboration (SMC) at CERN
provide a measurement of the polarization difference,
integrated over $x$ and weighted by $e_{q}^{2}/e^{2}$ \cite{SMC}:
\begin{equation}
I \equiv \int_{0}^{1} dx \sum_{f=u,d,s,c}\frac{e_{f}^{2}}{e^{2}}
\frac{1}{2}\left[\Delta{f}(x,Q^{2})
+\Delta{\bar{f}}(x,Q^{2})\right]
\end{equation}
\[=0.142\pm0.008\pm0.011,\]\noindent
where the momentum scale is $Q^{2}=10$ GeV$^{2}$.
The ppdf's used in our calculations have
\begin{equation}
I=0.138 {\rm ~for~} Q^{2}=10 {\rm ~GeV}^{2}~,
\end{equation}
\[I=0.163 {\rm ~for~} Q^{2}=(80 {\rm ~GeV})^{2}~,\]\noindent
where the $Q^{2}$-dependence is determined by renormalization group
equations.  For $W^{\pm}$ production the relevant momentum scale
is of order $M_{W}$.
The spin-dependence of the quark densities is rather small,
as indicated by the small value of $I$, so one question
that motivates our study is whether the cross-sections
for these $W^{\pm}$-production processes depend significantly on
the proton helicity.

In Section II we calculate the cross-section for the process
$p_{\lambda}\bar{p}\rightarrow W^{+}\gamma$,
where $\lambda=L$ or $R$ denotes a left-handed or right-handed proton.
This process is sensitive
to the $WW\gamma$ anomalous couplings.  In Section III we consider
the process $p_{\lambda}\bar{p}\rightarrow W^{\pm}~W^{\mp}$,
which is sensitive to both $WW\gamma$ and $WWZ$ anomalous couplings.
The purpose of these calculations is to explore whether polarization
of the protons can increase the sensitivity of
measurement of anomalous couplings.

\newpage
\section{$W^{+}\gamma$ production}

    The Feynman diagrams for the process
$p+\bar{p}\rightarrow{W^{+}+\gamma}$ are shown in Figure 2.1.
One diagram has a $WW\gamma$ vertex, so the cross-section
depends on the anomalous photon coupling parameters $\Delta\kappa_\gamma$
and $\lambda_\gamma$.  This process can be used to place limits on
the anomalous couplings; calculations of the unpolarized
cross-section with anomalous couplings were
described in Refs. \cite{BB} and \cite{BHO}.
For polarized scattering we expect
the cross-section for left-polarized protons
to be larger than for right-polarized protons,
because the produced $W^{+}$ line is always
connected to a left-handed quark line.
To investigate whether polarizing the proton beam would yield better
limits on  $\Delta\kappa_\gamma$ and $\lambda_\gamma$,
we have calculated the polarized and unpolarized cross-sections.
The results of this study are reported in this section.

\noindent{\bf a. Method of calculation}

    The cross-section for $p_{\lambda}+\bar{p}\rightarrow W^{+}+\gamma$,
where $\lambda$ is $L$ or $R$ for left-handed or right-handed protons,
and with subsequent decay $W^{+}\rightarrow \ell^{+}+\nu_{\ell}$, is
expressed as
\begin{equation}
\sigma(\lambda)=\int_{0}^{1}dx dx^{\prime}\left[
\hat{\sigma}_{LR}(xP_{1},x^{\prime}P_{2})u_{\pm}(x)d(x^{\prime})\right.
\end{equation}
\[\left.~~~~~~~~~~+\hat{\sigma}_{LR}(x^{\prime}P_{2},xP_{1})
\bar{d}_{\mp}(x)\bar{u}(x^{\prime})\right]
\]\noindent
where the notation is as follows:
The parton
cross-section $\hat{\sigma}_{LR}(p_{1},p_{2})$ is for the process
$u_{L}(p_{1})+\bar{d}_{R}(p_{2})\rightarrow W^{+}+\gamma$.
The upper sign on $u_{\pm}(x)$ and $\bar{d}_{\mp}(x)$ is for
$\lambda=L$ and the lower sign is for $\lambda=R$.
The first line in Eq.\ (10)
corresponds to $u,\bar{d}$ coming from $p,\bar{p}$, and the second line
corresponds to $\bar{d},u$ coming from $p,\bar{p}$, respectively.
$x$ and $x^{\prime}$ are the parton momentum fractions
in the proton and antiproton respectively.
The parton distribution functions are, for example,
\[u_{\pm}(x) = u {\rm ~quark~with~same/opposite~helicity~as~} p\]
\[d(x) = \bar{d} {\rm ~quark~in~unpolarized~} \bar{p}\]
\[\bar{d}_{\mp}(x) = \bar{d} {\rm ~quark~with~opposite/same~helicity~as~} p\]
\[\bar{u}(x) = u {\rm ~quark~in~unpolarized~} \bar{p}~.\]\noindent
We also add the contribution for the parton process $c+\bar{s}\rightarrow
W^{+}+\gamma$, which is, however, small.
(We ignore Cabibbo-Kobayashi-Maskawa mixing in this work.)
Finally, we add the cross-sections for two lepton
decay modes of the $W^{+}$; that is, $\ell$ can be
either $e$ or $\mu$.

The parton cross-section is calculated from helicity amplitudes for
the reaction.  In this reaction there is only one nonzero helicity
amplitude, with $\lambda_{u}=L$ and $\lambda_{\bar{d}}=R$,
because we approximate the quark masses as 0.
In our calculation we calculate the helicity amplitude
${\cal M}_{LR}$ numerically.
Then the parton cross-section is
\begin{equation}
\hat{\sigma}_{LR}=\int d\Phi |{\cal M}_{LR}|^{2}
\end{equation}\noindent
where $\int d\Phi$ indicates a phase space integral.

The phase space and $x,x^{\prime}$ integrations are performed
by a Monte-Carlo program, based on the Vegas Monte-Carlo
integration routine \cite{LePage}.
The style of the full Monte-Carlo program is the same as the
program PAPAGENO \cite{Hinchliffe}.

The kinematic cuts we impose on the final $\ell^{+}$ and $\gamma$ are:
\begin{equation}
{\rm rapidity~} |\eta_{\ell}|<3~,~~|\eta_{\gamma}|<3~,
\end{equation}
\begin{equation}
{\rm transverse~momentum~} p_{T\ell}>20 {\rm ~GeV}~,
{}~~p_{T\gamma}>20 {\rm ~GeV}~,
\end{equation}
\begin{equation}
\Delta{R}\equiv\sqrt{(\Delta\eta)^{2}+(\Delta\phi)^{2}}>0.7~,
\end{equation}\noindent
where $\Delta{R}$ is the separation of the $\ell^{+}$ and $\gamma$
in $\eta-\phi$ space.
The only cut on the neutrino is a transverse momentum cut
$p_{T\nu}>20$ GeV; that is, we require
\begin{equation}
E\!\!\!/_{T}>20 {\rm ~GeV}~.
\end{equation}

At the parton level, there is no background to this process
from other interactions, as long as we require the $W^{+}$
to decay to leptons.
There is an experimental backround due to confusion
between jets and photons in the detector \cite{BHO}.
However, we do not consider the experimental background here,
because our interest is to examine the effect of proton
polarization, compared with the unpolarized case.

\noindent{\bf b. Results}

Tables 2.1 and 2.2 show the results of our calculations: the
$W^{+}\gamma$ production
cross-section for polarized and unpolarized protons with
different values of anomalous couplings
$\Delta\kappa_{\gamma}$ and $\lambda_{\gamma}$.
The cross-section includes the branching ratio $2/9$
for $W^{+}\rightarrow \ell^{+}\nu_{\ell}$,
where the decay modes $\ell=e$ and $\ell=\mu$ are added.
(The branching ratio factor $2/9$ is included in all
cross-sections reported hereafter.)
The cross-section is smallest for $\Delta\kappa_{\gamma}=0$
and $\lambda_{\gamma}=0$, {\it i.e.} the Standard Model values.
The cross-section depends more strongly on
$\lambda_{\gamma}$ than on $\Delta\kappa_{\gamma}$.
Figures 2.2 and 2.3 show plots of the cross-section
{\it vs} $\Delta\kappa_{\gamma}$ assuming $\lambda_{\gamma}=0$,
and {\it vs} $\lambda_{\gamma}$ assuming $\Delta\kappa_{\gamma}=0$.
As expected, the cross-section is larger for left-handed protons;
$\sigma(L)$ is roughly 3 times $\sigma(R)$, and so roughly
1.5 times the unpolarized cross-section.
The unpolarized cross-section is by definition
equal to $\frac{1}{2}(\sigma(L)+\sigma(R))$.

Figures 2.2 and 2.3 show only the total cross-section
(for the cuts specified in Eqs.\ (12)-(15)).
Analysis of differential cross-sections, with respect to
relevant kinematic variables, may provide more precise
tests of anomalous couplings \cite{BHO}.
For example, Figure 2.4 shows the distribution of $p_{T\gamma}$
for polarized proton scattering,
with $\Delta\kappa_{\gamma}=0$ (solid line)
and $\Delta\kappa_{\gamma}=-1$ (dashed line).
The shapes of the distributions are similar for left- or right-handed
protons, but there is an overall difference of magnitude.

\noindent
{\bf{c. Limits on $\Delta\kappa_{\gamma}$ and $\lambda_{\gamma}$}}

Tables 2.1 and 2.2 are results from the Monte Carlo
calculations of cross-section {\it vs} anomalous couplings.
There is a fairly
large effect of proton polarization: $\sigma(L)$ is generally
about a factor of 3 larger than $\sigma(R)$.
But to see whether an experiment with polarized protons would
yield a significantly better measurement of the anomalous
couplings, we must estimate the experimental limit that could
be set on $\Delta\kappa_{\gamma}$ or $\lambda_{\gamma}$,
for a given integrated luminosity.

To estimate the experimental limit that could be placed
on an anomalous $WW\gamma$ coupling, we must estimate the
uncertainty in an experimental measurement of
the cross-section $\sigma$.
For this analysis we simply assume that the
standard deviation in the number of events $N$
is $\delta N=\sqrt{N}$,
{\it i.e.}\ that $N$ obeys Poisson statistics.
(This is an underestimate of the experimental uncertainty;
for instance, it does not take into account the experimental
background of jets misidentified as photons, {\it etc}.
For our purposes we do not include the efficiency of the detector.)
The measured cross-section would be
$\sigma=N/L$ where $L$ is the integrated luminosity.
The 3-sigma upper limit on $\sigma$
({\it i.e.}~99.7\% confidence level)
expected from Poisson statistics would be $(N+3\sqrt{N})/L$.
Thus the measurement would rule out a cross-section
larger than $\sigma+\delta\sigma_{3}$, where
\begin{equation}
\delta\sigma_{3}=3\sqrt{\frac{\sigma}{L}}~.
\end{equation}\noindent
To estimate the limit that could be placed on
$\Delta\kappa_{\gamma}$ or $\lambda_{\gamma}$
(assuming the anomalous couplings are in fact zero),
we compare the uncertainty $\delta\sigma_{3}$ to the variation
of the calculated $\sigma$ as a function of the anomalous coupling.
At the 3-sigma confidence level
$\Delta\kappa_{\gamma}$ would be in the range with
\begin{equation}
|\sigma(\Delta\kappa_{\gamma})-\sigma(\Delta\kappa_{\gamma}=0)|
<\delta\sigma_{3}~,
\end{equation}\noindent
and similarly for $\lambda_{\gamma}$.

We assume a Tevatron integrated luminosity equal to
1 fb$^{-1}$ or 10 fb$^{-1}$.
We assume the {\em same} integrated luminosity for
each polarization and for the unpolarized case.
Then we obtain the results in Table 2.3, for the
experimental limits that could be set on the two
anomalous couplings.
We emphasize that the numbers in Table 2.3 do not include
possible experimental uncertainties;
our purpose in estimating the limits that could be placed on
$\Delta\kappa_{\gamma}$ and $\lambda_{\gamma}$ is only to
compare the precision from polarized or unpolarized protons.

The left-polarized proton provides a better limit on
$\Delta\kappa_{\gamma}$ or $\lambda_{\gamma}$, because
it has a larger cross-section.
The improvement in precision from left-polarized protons,
compared to unpolarized, is not very great, because
$\sigma(L)$ is only about 1.5 times larger than
$\sigma(unpolarized)$, and because the precision
on $\sigma$ is only proportional to $\sigma^{1/2}$.
The result is that the
total cross-section for left-polarized proton
scattering can provide a better limit on $\Delta\kappa_{\gamma}$
or $\lambda_{\gamma}$ than that for unpolarized scattering,
better by about 10 to 20 \%.

Another way to consider the effect of proton helicity
is to calculate the left-right asymmetry, defined by
\begin{equation}
{\cal A}=\frac{\sigma(L)-\sigma(R)}{\sigma(L)+\sigma(R)}~.
\end{equation}\noindent
An asymmetry measurement may be accurate experimentally
because systematic errors cancel in the ratio.
The possible range of ${\cal A}$ is $-1\le{\cal A}\le 1$, and
${\cal A}$ is equal to $0$ in a left-right symmetric world.
For the {\em parton} process
$q_{\lambda}\bar{q}^{\prime}\rightarrow W^{+}\gamma$,
we have ${\cal A}=1$ because only left-handed quarks contribute.
For the {\em proton} process, the asymmetry is reported
in Tables 2.1 and 2.2; we find ${\cal A}\approx 0.5$.
However, ${\cal A}$ depends only weakly on
$\Delta\kappa_{\gamma}$ and $\lambda_{\gamma}$:
${\cal A}$ varies by ${\cal O}(10\%)$ over the range of
anomalous couplings considered.
Thus a measurement of ${\cal A}$ to determine
$\Delta\kappa_{\gamma}$ or $\lambda_{\gamma}$
would require high statistics.

We have only considered the process
$p_{\lambda}\bar{p}\rightarrow W^{+}\gamma$, and not $W^{-}\gamma$,
because the former is more sensitive to the proton helicity.
A $W^{+}$ comes from a $u$ quark, whereas a $W^{-}$ comes from
a $d$ quark.
The helicity dependence is stronger for $u$ than $d$ in a proton,
as seen in Figure 1.1.

\newpage
\section{$W^{+}W^{-}$ production}

The process $p\bar{p}{\rightarrow}W^{+}W^{-}$
provides a way to test the Standard Model $WWV$
vertices for both $V=\gamma$ and $V=Z^{0}$.
We consider the {\em doubly} electroweak process
\begin{equation}
p(\lambda)+\bar{p}\rightarrow
W^{+}\left(\rightarrow \bar{\ell}+\nu_{\ell}\right)
+ W^{-}\left(\rightarrow d+\bar{u}\right)~,
\end{equation}\noindent
and also the process in which the $W^{-}$ decays leptonically
while the $W^{+}$ decays to 2 jets.
The complete set of Feynman diagrams with the final state
$\bar{\ell}\nu d \bar{u}$ includes diagrams that do not
have the form of $W^{+}W^{-}$ production.
Figure 3.1 shows the complete set of Feynman diagrams for the
parton process
\begin{equation}
u+\bar{u}\rightarrow W^{+}+d+\bar{u}~,
{\rm ~~with~} W^{+}\rightarrow \bar{\ell}+\nu_{\ell}~;
\end{equation}\noindent
there is similarly a set of diagrams for the process
\begin{equation}
d+\bar{d}\rightarrow W^{+}+d+\bar{u}~,
{\rm ~~with~} W^{+}\rightarrow \bar{\ell}+\nu_{\ell}~.
\end{equation}\noindent
Also, there are similar diagrams for production of
$W^{-}+u+\bar{d}$, with $W^{-}\rightarrow \ell+\bar{\nu}_{\ell}$.
(The complete set of diagrams for the final state
$\bar{\ell}\nu d\bar{u}$,
or $\ell\bar{\nu}u\bar{d}$,
includes additional diagrams in which the leptons are not
decay products of a single narrow $W^{\pm}$.)
These parton processes involve the $WW\gamma$ and $WWZ$
vertices in some Feynman diagrams, so the cross-sections
depend on anomalous $WWV$
couplings, {\it i.e.} the parameters
$\Delta\kappa_\gamma, \Delta\kappa_Z, \lambda_\gamma, \lambda_Z$.
In this section we study the cross-section as a function of
these non-Standard parameters, for polarized protons.
The purpose is again to see whether an experiment with
a polarized proton beam would yield a stronger
test of the electroweak triple-boson vertices.

Some, though not all, of the Feynman diagrams in Fig.\ 3.1
have the form of $W^{+}W^{-}$ production, followed by decays
of the $W$'s, one leptonically and the other into two jets.
At 2 TeV center-of-mass energy, the cross-section is dominated
by the $W^{+}W^{-}$ production.
Therefore, as explained further below,
we approximate the cross-section by $W^{+}W^{-}$ production.
The cross-sections for $\ell \bar{\nu} u \bar{d}$
and $\bar{\ell} \nu \bar{u} d$ final states are equal in this
approximation, because either $W$ is equally likely to decay
leptonically.

A related process is $W^{\pm}Z^{0}$ production, where the $Z^{0}$
decays to 2 jets, {\it e.g.,}
\begin{equation}
u+\bar{d}\rightarrow W^{+}\left(\rightarrow \ell^{+}+\nu_{\ell}\right)
+Z^{0}\left(\rightarrow q+\bar{q}\right)~.
\end{equation}
In our calculations we ignore the $W^{\pm}Z^{0}$ production.
In a theoretical calculation the $W^{+}W^{-}$ production is
distinguishable from the $W^{\pm}Z^{0}$ production.
However, in an experiment these two processes are
tangled together, because they are both observed
as $W^{\pm}+2~jets$.
In our $W^{+}W^{-}$ calculation we impose a kinematic
cut on the invariant mass $M_{2j}$ of the 2 jets,
making $M_{2j}$ approximately equal to the $W^{\pm}$ mass;
specifically we take $M_{2j}$ between 70 and 90 GeV.
(This cut reduces the QCD background of $W^{\pm} + 2~jets$.)
But even with this cut on $M_{2j}$
there would still be an overlap between $W^{+}W^{-}$ production
and $W^{\pm}Z^{0}$ production.
The purpose of our calculation is a theoretical study
of the effect of proton polarization on the cross-section.
A complete analysis of experimental data would need to
include both the $W^{+}W^{-}$ and $W^{\pm}Z^{0}$
processes together.

The method of calculation is similar to Sec.\ 2, but with
some differences.  In Sec.\ 2 only one parton helicity
combination contributes, $u(L)\bar{d}(R)$.
Here two helicity combinations contribute,
for example $u(L)\bar{u}(R)$ and $u(R)\bar{u}(L)$, and we
include both.
In fact, for Standard couplings the contribution from
$u(R)\bar{u}(L)$ is very small
compared to $u(L)\bar{u}(R)$,
because of interference between Feynman diagrams,
so the parton-level process still depends
strongly on quark helicities.
A more important difference is that in Sec. II there was
no parton-level background process, whereas here we have
a large background from electroweak+QCD processes.

\noindent{\bf a. Background, Approximations, and Cuts}

The parton-level background to this process is the production of
$W^{\pm} + 2~jets$ by processes with one electroweak vertex and one
QCD vertex.
These processes do not interfere quantum mechanically with
our doubly electroweak signal process, because they have a
different color structure;
for example, the $q\bar{q}g$ vertex
is color octet, where $g$ denotes the gluon,
whereas $q\bar{q}\gamma$ or $q\bar{q} Z^{0}$
vertices are color singlet.  However the final states are
indistinguishable experimentally, so the doubly electroweak
reaction is hidden in a background of electroweak-QCD reactions.

We have calculated the background cross-section from
the helicity amplitudes for a complete set of $W^{\pm}+2~jets$
processes \cite{HZ}, with the polarized parton distribution
functions described in Sec.\ 1.
The background reactions should depend
strongly on the helicity of the proton:
The produced $W^{\pm}$ must connect to a left-handed quark
or right-handed antiquark for massless quarks,
by the V$-$A coupling;
the density of quarks of given helicity
depends on the helicity of the proton.
By contrast, for the signal reaction the diagrams
{\em with anomalous $WW\gamma$ or $WWZ$ vertices} do not
require any specific helicities of the incoming quarks.
Thus the signal reaction will have a different dependence
on proton helicity than the background.
The important question is
whether the signal-to-background ratio
in the measurement of anomalous $WWV$ couplings
is better for left- or right-polarized protons.

To reduce the background we impose a cut on the invariant
mass $M_{2j}$ of the 2 jets, putting $M_{2j}$ approximately
equal to $M_{W}$.
The solid curve in Figure 3.2 shows the distribution
of the invariant mass of the two jets produced by the complete
doubly electroweak process with final state
$q\bar{q}^{\prime}W^{+}$,
with $W^{+}\rightarrow \bar{\ell}\nu_{\ell}$,
for $\sqrt{s}=2$ TeV.
The $q\bar{q}^{\prime}$ invariant mass is peaked
at the $W^{+}$ mass.
Because the cross-section is dominated by the
$W^{+}$ resonance, we
approximate the calculation by keeping only
the Feynman diagrams that produce a $W^{+}W^{-}$ pair,
which is an accurate simplifying approximation.
Furthermore, we {\em require} the $q\bar{q}^{\prime}$
invariant mass to be approximately equal to the $W^{\pm}$ mass:
we calculate the cross-section only for events with
$M_{q\bar{q}^{\prime}}$ between 70 and 90 GeV.
The dotted curve in Figure 3.2 shows the 2-jet mass distribution
when we neglect all but the $W^{+}W^{-}$ pair production diagrams
and also require the 2-jet mass to be between 70 and 90 GeV.
With this $M_{2j}$ cut, the complete
doubly electroweak process is practically
the same as production of $W^{+}W^{-}$ followed by
leptonic decay of $W^{\pm}$ and quark-antiquark decay
of $W^{\mp}$.
For comparison, Figure 3.3 shows the 2-jet mass distribution
of the background processes with the kinematic cuts listed
in Eqs.\ (23)-(26) below;
the cut $70 {\rm ~GeV} < M_{q\bar{q}^{\prime}} < 90 {\rm ~GeV}$
reduces the total background significantly,
because there is no resonant effect in the background processes.

In addition to the $M_{2j}$-cut just described,
we impose the following kinematic cuts on all
the final-state particles except the neutrino:
\begin{equation}
{\rm rapidity~} |\eta|<3,
\end{equation}
\begin{equation}
{\rm transverse~momentum~} p_{T}>15 {\rm ~GeV},
\end{equation}
\begin{equation}
\Delta{R}\equiv\sqrt{({\Delta\eta})^2+({\Delta\phi})^2}>0.7~,
\end{equation}
where $\Delta{R}$ is the separation of any
pair of final particles not including the neutrino.
In the case of the neutrino, we impose only a transverse
momentum cut, $p_{T\nu}>15$ GeV; {\it i.e.,}
\begin{equation}
E\!\!\!/_{T} > 15 {\rm ~GeV.}
\end{equation}\noindent
Figure 3.4 is another comparison between the
complete doubly electroweak calculation, shown as the solid line,
and the simplified approximation ($W^{+}W^{-}$ production with
subsequent $W^{\pm}$ decays), shown as the dashed line.
Figure 3.4 compares the $\sqrt{\hat{s}}$ distributions,
where $\sqrt{\hat{s}}$
is the center of mass energy of the parton-level process.
Again, the two calculations are practically equal.

We also consider, separately, a cut requiring large $\sqrt{\hat{s}}$,
specifically $\sqrt{\hat{s}}>340 {\rm ~GeV}$,
The variable $\sqrt{\hat{s}}$ is important because the effect
of anomalous coupling increases with $\sqrt{\hat{s}}$.
Figure 3.5 compares signal and background as a function of
$\sqrt{\hat{s}}$.
This figure shows why large-$\sqrt{\hat{s}}$ is interesting:
the signal-to-background ratio is larger,
and the dependence on $\Delta\kappa$ and
$\lambda$ is stronger, for large $\sqrt{\hat{s}}$.
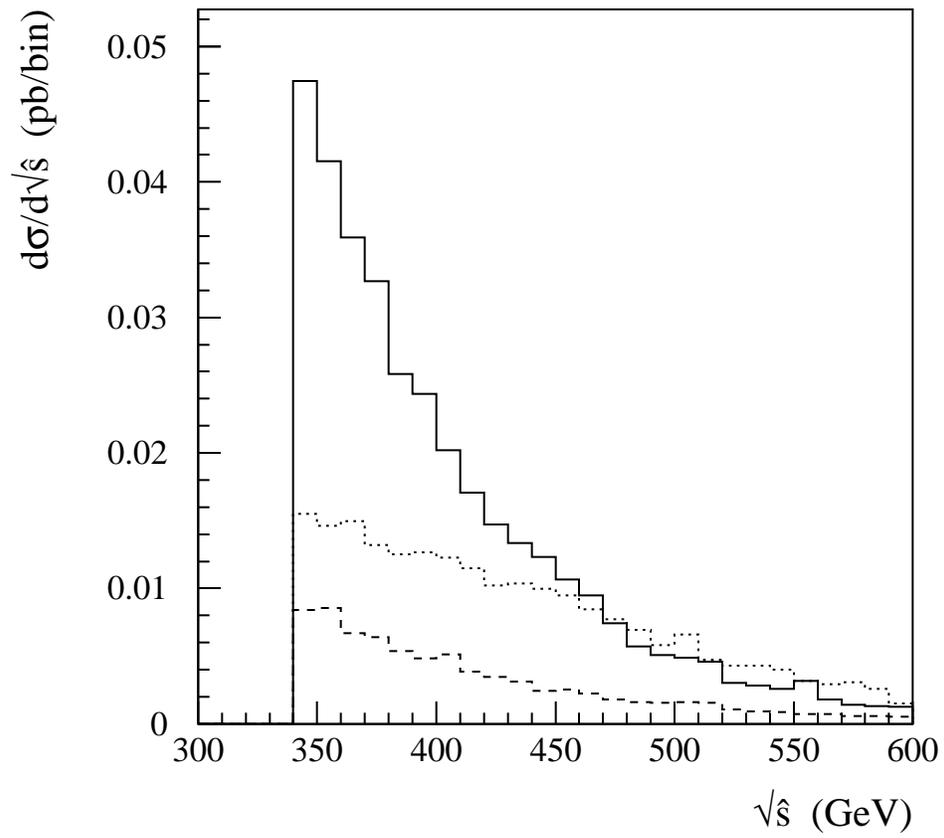
Figure 3.6 compares signal and background for
$\sqrt{\hat{s}}>340 {\rm ~GeV}$.
On the other hand,
this energy cut reduces drastically the
total number of events, so it becomes a question of
detailed calculation to see whether it is a real advantage,
given the available luminosity.
Our calculation is for $\sqrt{s}=2$ TeV.
A $p\bar{p}$ collider with higher center-of-mass energy
would produce more events in the interesting region of
phase space with large $\sqrt{\hat{s}}$,
and provide stronger tests of the anomalous $WWV$ couplings.

It is difficult to determine $\sqrt{\hat{s}}$ accurately
in an experimental event, because that requires measurement
of jet momenta and missing neutrino momentum.
The $z$-component of the neutrino momentum $p_{\nu z}$
can be obtained up to a two-fold ambiguity by solving
the mass contraint for the $W$-boson,
$M_{W}^{2}=(p_{\ell}+p_{\nu})^{2}$.
One way to choose $p_{\nu z}$ is to select the
solution with smaller absolute value, because a hard
scattering process tends to produce final products in the
central radidity region with large transverse momenta.
Given $p_{\nu z}$ it is straightforward to calculate
$\sqrt{\hat{s}}$ from the 4-momenta of $\ell,\nu,$
and the 2 jets.
Alternatively, another way to select large $\sqrt{\hat{s}}$
is to require large transverse mass of the final state
of the hard scattering process \cite{WLWL};
however, we have not pursued that approach in this work.

\noindent{\bf b. Results}

For the cross-section calculations that follow,
we set $Q=\sqrt{\hat{s}}$, where $Q$ is the momentum scale
used in the parton distribution functions .
The uncertainties we quote are only the statistical
Monte-Carlo uncertainties.
The theoretical uncertainty due
the choice of $Q$ scale is discussed briefly later.
The cross-section includes the branching ratio 2/9 for
leptonic decay of one $W$
(into either electron or muon plus neutrino).
The cross-section also includes the branching ratio 6/9 for
2-jet decay of the other $W$,
{\it e.g.} $W^{+}\rightarrow u\bar{d}$ and
$W^{+}\rightarrow c\bar{s}$.
(Again we ignore CKM mixing.)
The overall branching ratio
$(2/9)\times(6/9)=4/27$ is always included in the
cross-sections reported hereafter for $W^{+}W^{-}$ production.
Note that we consider separately the rates for
$\ell^{+}+jets$ and $\ell^{-}+jets$.

Figures 3.7 to 3.9 show the results as plots of
signal cross-section {\it vs} anomalous coupling parameters.
Since our signal process involves the creation of
a $W^{+}W^{-}$ pair with subsequent decays of the
$W^{+}$ and $W^{-}$,
the signal cross-section is the same regardless
of which $W$ decays to leptons, in the narrow-width approximation.
We need to compare the signal cross-section to the
QCD background cross-section.
We may consider the background for either $W^{+}$ or $W^{-}$
production, and then choose the $W$ charge that has the smaller
background cross-section.
For {\em polarized} protons, the background cross-sections
for $W^{+}$ or $W^{-}$ production are different.
For instance, for a left-handed proton the background process
$p(L)+\bar{p}\rightarrow W^{+}+2~jets$ has a larger rate than
$p(L)+\bar{p}\rightarrow W^{-}+2~jets$ , because the probability
of finding $u(L)$ (which produces $W^{+}$) inside $p(L)$ is larger
than $d(L)$ (which produces $W^{-}$),
as implied by Fig.\ 1.1;
$u(R)$ and $d(R)$ do not contribute to the constituent
cross-section of the $W^{\pm}+2~jets$ background process because
the weak charged current is left-handed.
Therefore it is advantageous experimentally to observe the
$W^{-}(\rightarrow \ell\bar{\nu})+2~jets$
mode for a left-polarized proton beam, and similarly
the $W^{+}(\rightarrow \bar{\ell}\nu)+2~jets$
for a right-polarized proton beam.
(For {\em unpolarized} scattering the production rates
of $W^{+}$ and $W^{-}$ are equal by CP invariance;
but for scattering of polarized protons on unpolarized antiprotons,
CP invariance does not apply.)
Since the signal process is symmetric with respect to
the charge of the $W$ that decays leptonically, the
process with the smaller background has a better
signal-to-background ratio.
Hereafter we apply this strategy when comparing signal
and background rates in the Tables.

The program we have used to calculate the electroweak+QCD background,
which comes from many processes \cite{HZ},
is set up to calculate the cross-section for
production of $W^{-}+2~jets$.
To find the cross-section
for production of $W^{+}+2~jets$, we calculate the rate
for $W^{-}$ production with polarized {\em anti-protons},
which is equal to the rate we want by CP invariance:

\begin{equation}
\sigma(p+\bar{p}(L)\rightarrow W^{-}+2 {\it ~jets})
=\sigma(p(R)+\bar{p}\rightarrow W^{+}+2 {\it ~jets})
\end{equation}
\begin{equation}
\sigma(p+\bar{p}(R)\rightarrow W^{-}+2 {\it ~jets})
=\sigma(p(L)+\bar{p}\rightarrow W^{+}+2 {\it ~jets})
\end{equation}

Tables 3.1 to 3.5 give the calculated cross-sections.
Table 3.5 lists the background cross-sections for different
proton helicities.
As expected, $\sigma(L)$ is larger than $\sigma(R)$
for $W^{+}$ production: the $W^{+}$ must come from a $u(L)$,
and the density of $u(L)$ is larger in $p(L)$ than in $p(R)$.
On the other hand, $\sigma(L)$ is {\em smaller} than $\sigma(R)$
for $W^{-}$ production: the $W^{-}$ must come from a $d(L)$,
and the density of $d(L)$ is
{\em smaller} in $p(L)$ than in $p(R)$,
because $\Delta d(x)$ is mostly negative, as shown in Fig.\ 1.1.

Table 3.5 lists cross-sections calculated
using the polarized parton distribution functions of Sec.\ 1,
and also, for comparison, unpolarized cross-sections calculated
using CTEQ2 parton distribution functions \cite{pdfcteq2}.
We have used the leading-order set
CTEQ2L for our studies.
The CTEQ2 results are consistent with
the average of the two proton helicities,
within the uncertainty of the Monte Carlo calculations.

\noindent
{\bf c. Limits on $\Delta\kappa_{\gamma}$, $\Delta\kappa_{Z}$,
            $\lambda_{\gamma}$, and $\lambda_{Z}$}

We estimate limits on anomalous couplings that could be set by
experiments at $\sqrt{s}=2$ TeV.
Since there is a large background for this process, the
limits depend on the background
cross-section $\sigma_{B}$.  Again, as in our analysis of
the $W^{+}\gamma$ process,
we assume that in an experiment with $N$ events
the standard deviation of $N$ is $\delta{N}=\sqrt{N}$.
At the three-sigma confidence level, the uncertainty
of $\sigma$ is
\begin{equation}
\delta\sigma_{3}=3\sqrt{\frac{\sigma}{L}}~,
\end{equation}\noindent
which is the same as Eq.\ (16), except that here $\sigma$
is the sum of signal and background cross-sections.
We consider integrated luminosity
$L=1$ fb$^{-1}$ and $L=10$ fb$^{-1}$.

To calculate the limit that could be set
on an anomalous coupling parameter, {\it e.g.}
$\Delta\kappa_{\gamma}$,
assuming the actual value of the parameter is zero,
we compare the statistical
uncertainty $\delta\sigma_{3}$ to the variation of
the calculated cross-section as a function of
$\Delta\kappa_{\gamma}$.
At the three-sigma confidence level, $\Delta\kappa_{\gamma}$
is in the range with
\begin{equation}
|\sigma(\Delta\kappa_{\gamma})-\sigma(\Delta\kappa_{\gamma}=0)|
<\delta\sigma_{3}~.
\end{equation}
For example, Figure 3.7c shows $\sigma$ {\it vs} $\Delta\kappa$
for unpolarized scattering.
Three possibilities are shown: ({\it i})
the effect of $\Delta\kappa_{\gamma}$ with $\Delta\kappa_{Z}=0$,
({\it ii}) the effect of $\Delta\kappa_{Z}$ with
$\Delta\kappa_{\gamma}=0$, and ({\it iii}) the effect
if $\Delta\kappa_{\gamma}=\Delta\kappa_{Z}$.
The background cross-section is given in Table 3.5.
Then Tables 3.6 and 3.7 list the limits that could be set
on $\Delta\kappa$ (and also $\lambda$) for these three
cases.
The estimated limits in Table 3.6 are from
events with arbitrary $\sqrt{\hat{s}}$, and the limits
in Table 3.7 are from events with $\sqrt{\hat{s}}>340$ GeV.

Our purpose is to determine whether measurment of the
cross-section with polarized protons leads to stronger
limits on anomalous couplings than with unpolarized protons.
Tables 3.6 and 3.7 show that the limit is stronger with
left-polarized protons, assuming the same integrated
luminosity as for unpolarized scattering.
The constraints on either $\Delta\kappa_{\gamma}$
or $\lambda_{\gamma}$ alone from the $W^{+}W^{-}$ channel are not
as strong as those obtained from studying the $W^{+}\gamma$ channel.
The constraints on $\Delta\kappa_{Z}$ and $\lambda_{Z}$
are about a factor of 2 better than those on
$\Delta\kappa_{\gamma}$ and $\lambda_{\gamma}$
from the $W^{+}W^{-}$ channel.
For models with a linear SU(2) symmetry, such that
$\kappa_{\gamma}=\kappa_{Z}$ and $\lambda_{\gamma}=\lambda_{Z}$,
the constraint on $\Delta\kappa_{\gamma}$ is about a factor
of 2 better than from the $W^{+}\gamma$ channel,
while the constraint on $\lambda_{\gamma}$ is about the
same as from the $W^{+}\gamma$ channel, after selecting
the large $\sqrt{\hat{s}}$ region.
In general, selecting large $\sqrt{\hat{s}}$
($>340$ GeV) improves the significance of signal
to background by about a factor of 2.

\noindent
{\bf d. Interpretation of the results}

The limits described above show how this process,
$W^{+}W^{-}$ production with polarized protons,
tests the $WWV$ vertex.
The estimated limits on anomalous couplings include
only simple statistical uncertainty,
based on Poisson statistics ($\delta{N}=\sqrt{N}$);
they do not include experimental uncertainty due to
detector inefficiency or theoretical uncertainty of
parton distribution functions or $Q$ scale.

For example, the background calculation depends on
the choice of parton momentum scale $Q$.
Table 3.8 shows how the background cross-section,
for unpolarized scattering, depends on the choice of $Q$.
We used $Q=\sqrt{\hat{s}}$ in our calculations.
$Q=2M_W$ or $Q=\sqrt{\hat{s}}/2$ would also be reasonable choices.
The cross-sections for the three choices differ by about 20\%.
This theoretical uncertainty due to dependence on scale choice is
larger than the statistical uncertainty analysed above.
It is thus difficult to extract the signal cross-section
unless the uncertainty in the background cross-section is
reduced, by including higher-order QCD effects to reduce the
dependence on $Q$ scale.
On the other hand,
given large luminosity the details of the QCD background processes,
{\it e.g.} the shapes of $\sqrt{\hat{s}}$ distributions, can be measured
directly from data to discriminate between different
theoretical predictions from different scale choices.
Therefore we anticipate that background rates will
eventually be better known, and allow us to extract signal rates.

Our purpose in this work is to study the
effect of proton polarization.
For this study we have analysed only the {\em total
cross-section} as a function of $\Delta{\kappa}$ and $\lambda$.
Stronger constraints on $\Delta{\kappa}$ and $\lambda$
may be set by analysing differential cross-sections,
with respect to variables for which the distribution
of events is sensitive to $\Delta{\kappa}$ or $\lambda$
\cite{BB,BHO}.
It would be interesting to find kinematic variables
{\em dependent on proton helicity} for which the
distribution of events is sensitive to $\Delta{\kappa}$
or $\lambda$, to further test the triple-boson couplings
with polarized proton scattering.

\newpage
\section{Discussion and Conclusions}

We have studied the potential of a polarized proton beam at the
Tevatron collider for measuring the tri-boson couplings
$WW\gamma$ and $WWZ$.
Because the polarized parton distribution functions in the
relevant kinematic region ({\it i.e.}, $x$-values)
are not yet precise enough to give definite detailed
predictions about the rates of the signal and the backgrounds,
we have concentrated on  the comparisons between the total event
rates from a polarized- and an unpolarized- proton beam.
As summarized in Tables 2.3, 3.6 and 3.7, we found
that with a polarized proton beam the limits on
non-standard parameters
$\Delta \kappa_\gamma$, $\lambda_\gamma$,
$\Delta \kappa_Z$ and $\lambda_Z$
are somewhat improved compared to those obtained from
an unpolarized proton beam.
We anticipate that these results can be further improved by
studying detailed distributions of relevant kinematic variables.
Generally a factor of 2 improvement in measuring these
non-standard parameters is expected, after selecting kinematic
regions where the signal becomes more important,
as illustrated in Tables 3.6 and 3.7.

Reliable predictions of distributions of kinematic variables
will require more precise polarized parton distribution functions.
Measurements of polarized proton-antiproton reactions, such as
single-$W^{\pm}$ production with a polarized proton beam, will
yield new information on spin-dependent parton distributions.
Thus the use of polarized scattering to test fundamental physics,
and the determination of the spin structure of the proton,
would proceed together.

One interesting feature that we found about the polarized
collider program is that it is possible to select the
polarization state of the proton beam to enhance the ratio of
signal to background for a specific charge mode of the final state.
This was demonstrated in studying the process
$p {\bar p} \rightarrow W^+ W^- \rightarrow
\ell^\pm\,{\nu\!\!\!\!^{^{(-)}}\!}_{\ell}\, + \,\, 2 \,\, jets$.
For the final state with a positive charged lepton ($\ell^+$),
one can select the polarization of the proton beam to
be right-handed to improve the signal-to-background
ratio because the QCD background process
$p(R) {\bar p} \rightarrow
W^+ ( \rightarrow \ell^+ \nu_\ell) + \,\, 2 \,\, jets$
has a smaller rate than
$p(L) {\bar p} \rightarrow
W^+ ( \rightarrow \ell^+ \nu_\ell) + \,\, 2 \,\, jets$.
The signal rate, in contrast, is independent of the charge mode of
the isolated lepton from the $W$-boson decay because either $W^+$
or $W^-$ can decay into the charged lepton.
We expect that similar tricks can be applied to other
polarized physics measurements,
such as the $lepton+jet$ mode of the $t\bar{t}$ pair production
from $q\bar{q}$ and $gg$ fusion processes.

\section*{\normalsize Acknowledgements}

We thank Ray Brock and Harry Weerts for discussions on
polarized proton scattering.  We especially thank Glenn Ladinsky
for providing us with his program to calculate polarized
parton distribution functions.
The work of C.--P. Y. was supported by NSF grant No.
PHY-9309902.
\newpage

\newpage

\section*{Tables}
\noindent
Table 2.1.
Cross-section for the process $p\bar{p}\rightarrow W^{+}\gamma$
with polarized protons, for different values of the anomalous coupling
$\Delta\kappa_{\gamma}$, assuming $\lambda_{\gamma}=0$.
Cross-sections are in pb.
The branching ratio $2/9$ for $W^{+}\rightarrow e^{+}\nu_{e}$
or $\mu^{+}\nu_{\mu}$ is included.
The unpolarized case was calculated separately using
CTEQ2 parton distribution functions, for comparison.
The asymmetry ${\cal A}$, defined in Eq.~(18), is calculated by fitting
the data to a parabola.

\begin{center}
\begin{tabular}{l l l l l}\hline\hline
\multicolumn{5}{l}{\sl $p_{\lambda}\bar{p}\rightarrow W^{+}\gamma$
cross-sections in pb}\\ \hline
$\Delta\kappa_{\gamma}$ & p(R) & p(L) & Unpol & ${\cal A}$ \\ \hline
  2.0 & 0.49 $\pm$ 0.01 & 1.44 $\pm$ 0.01 & 0.98 $\pm$ 0.01 & 0.514\\
  1.5 & 0.39 $\pm$ 0.01 & 1.08 $\pm$ 0.01 & 0.75 $\pm$ 0.01 & 0.504\\
  1.0 & 0.29 $\pm$ 0.01 & 0.86 $\pm$ 0.01 & 0.59 $\pm$ 0.01 & 0.491\\
  0.5 & 0.25 $\pm$ 0.01 & 0.72 $\pm$ 0.01 & 0.51 $\pm$ 0.01 & 0.480\\
  0.0 & 0.25 $\pm$ 0.01 & 0.69 $\pm$ 0.01 & 0.47 $\pm$ 0.01 & 0.475\\
 -0.5 & 0.27 $\pm$ 0.01 & 0.75 $\pm$ 0.01 & 0.54 $\pm$ 0.01 & 0.480\\
 -1.0 & 0.32 $\pm$ 0.01 & 0.94 $\pm$ 0.01 & 0.63 $\pm$ 0.01 & 0.491\\
 -1.5 & 0.41 $\pm$ 0.01 & 1.23 $\pm$ 0.01 & 0.83 $\pm$ 0.01 & 0.504\\
 -2.0 & 0.53 $\pm$ 0.02 & 1.57 $\pm$ 0.02 & 1.09 $\pm$ 0.02 & 0.514\\
\hline\hline
\end{tabular}
\end{center}
\newpage

\noindent
Table 2.2.
Cross-section for the process $p\bar{p}\rightarrow W^{+}\gamma$
with polarized protons, for different values of the anomalous coupling
$\lambda_{\gamma}$, assuming $\Delta\kappa_{\gamma}=0$.
Cross-sections are in pb.
The unpolarized case was calculated separately using
CTEQ2 parton distribution functions, for comparison.
The asymmetry ${\cal A}$, defined in Eq.\ (18), is calculated by fitting
the data to a parabola.

\begin{center}
\begin{tabular}{l l l l l}\hline\hline
\multicolumn{5}{l}{\sl $p_{\lambda}\bar{p}\rightarrow W^{+}\gamma$
cross-sections in pb.}\\ \hline
$\lambda_{\gamma}$ & p(R) & p(L) & Unpol & ${\cal A}$ \\ \hline
  2.0 & 1.73 $\pm$ 0.03 & 7.47 $\pm$ 0.03 & 4.55 $\pm$ 0.03 & 0.613\\
  1.5 & 1.11 $\pm$ 0.02 & 4.47 $\pm$ 0.02 & 2.72 $\pm$ 0.02 & 0.602\\
  1.0 & 0.62 $\pm$ 0.01 & 2.32 $\pm$ 0.01 & 1.50 $\pm$ 0.01 & 0.580\\
  0.5 & 0.35 $\pm$ 0.03 & 1.14 $\pm$ 0.03 & 0.75 $\pm$ 0.03 & 0.530\\
  0.0 & 0.24 $\pm$ 0.01 & 0.68 $\pm$ 0.01 & 0.49 $\pm$ 0.01 & 0.478\\
 -0.5 & 0.34 $\pm$ 0.01 & 1.05 $\pm$ 0.01 & 0.71 $\pm$ 0.01 & 0.530\\
 -1.0 & 0.57 $\pm$ 0.01 & 2.19 $\pm$ 0.01 & 1.39 $\pm$ 0.01 & 0.580\\
 -1.5 & 0.99 $\pm$ 0.03 & 4.13 $\pm$ 0.03 & 2.59 $\pm$ 0.03 & 0.602\\
 -2.0 & 1.64 $\pm$ 0.03 & 6.87 $\pm$ 0.03 & 4.16 $\pm$ 0.03 & 0.613\\
\hline\hline
\end{tabular}
\end{center}
\newpage

\noindent
Table 2.3.
Limits on non-Standard couplings, from
$p\bar{p}\rightarrow W^{+}\gamma$ with polarized and
unpolarized protons.
The upper number is for 1 fb$^{-1}$ integrated luminosity,
and the lower number (in parentheses) is for 10 fb$^{-1}$.

\begin{center}
\begin{tabular}{l l l l} \hline\hline
Parameter & \multicolumn{3}{c} {\rm Limits}  \\ \hline
      & $p(R)$ & $p(L)$ & Unpol \\ \hline
$\Delta\kappa_\gamma$ &  $\pm0.89$  &  $\pm0.62$  &  $\pm0.70$ \\
                      & ($\pm0.50$) & ($\pm0.35$) & ($\pm0.39$)\\ \hline
$\lambda_\gamma$      &  $\pm0.36$  &  $\pm0.22$  &  $\pm0.26$ \\
                      & ($\pm0.20$) & ($\pm0.13$) & ($\pm0.14$)\\
\hline\hline
\end{tabular}
\end{center}
\newpage

\noindent
Table 3.1.
Purely electroweak cross-sections, in pb, for
$p(R)\bar{p}\rightarrow W^{+}+2~jets$, with $W^{+}\rightarrow\bar{\ell}\nu$
where $\ell=e$ or $\mu$;
the proton is right-handed.
The cross-section for $p(R)\bar{p}\rightarrow W^{-}+2~jets$
is the same.
The branching ratio $(2/9)\times(6/9)=4/27$ is included.

\begin{center}
\begin{tabular}{l | l l l l l l}\hline\hline
\multicolumn{7}{l}{\sl Electroweak cross-sections in pb;
 right-polarized proton.}\\ \hline
\multicolumn {1}{l|} { Value} &
\multicolumn {6} {c} {Varied quantity}\\ \hline
 & $\Delta\kappa_\gamma$ & $\Delta\kappa_Z$ &
 $\Delta\kappa_{\gamma}=\Delta\kappa_Z$ & $\lambda_\gamma$ &
 $\lambda_Z$ & $\lambda_{\gamma}=\lambda_Z$ \\ \hline
    1.00 &
       0.604$\pm$.012  &   1.120$\pm$.020  &   1.064$\pm$.020  &
       0.740$\pm$.020  &   1.344$\pm$.028  &   1.444$\pm$.032  \\
    0.75 &
       0.534$\pm$.012  &   0.774$\pm$.016  &   0.816$\pm$.016  &
       0.602$\pm$.014  &   0.952$\pm$.018  &   1.004$\pm$.020  \\
    0.50 &
       0.496$\pm$.012  &   0.600$\pm$.012  &   0.588$\pm$.012  &
       0.524$\pm$.012  &   0.688$\pm$.016  &   0.712$\pm$.016  \\
    0.25 &
       0.456$\pm$.012  &   0.484$\pm$.012  &   0.484$\pm$.012  &
       0.472$\pm$.012  &   0.520$\pm$.012  &   0.520$\pm$.012  \\
    0.00 &
       0.456$\pm$.012  &   0.456$\pm$.012  &   0.456$\pm$.012  &
       0.456$\pm$.012  &   0.456$\pm$.012  &   0.456$\pm$.012  \\
   -0.25 &
       0.484$\pm$.012  &   0.508$\pm$.012  &   0.516$\pm$.012  &
       0.480$\pm$.012  &   0.504$\pm$.012  &   0.516$\pm$.012  \\
   -0.50 &
       0.540$\pm$.012  &   0.628$\pm$.012  &   0.676$\pm$.012  &
       0.548$\pm$.012  &   0.664$\pm$.016  &   0.708$\pm$.016  \\
   -0.75 &
       0.620$\pm$.012  &   0.822$\pm$.014  &   0.908$\pm$.016  &
       0.642$\pm$.016  &   0.920$\pm$.018  &   1.000$\pm$.020  \\
   -1.00 &
       0.716$\pm$.016  &   1.172$\pm$.024  &   1.356$\pm$.028  &
       0.768$\pm$.016  &   1.288$\pm$.024  &   1.444$\pm$.028  \\
\hline\hline
\end{tabular}
\end{center}
\newpage
\noindent
Table 3.2.
Purely electroweak cross-sections, in pb, for
$p(L)\bar{p}\rightarrow W^{+}+2~jets$,
with $W^{+}\rightarrow\bar{\ell}\nu$
where $\ell=e$ or $\mu$;
the proton is left-handed.
The cross-section for $p(L)\bar{p}\rightarrow W^{-}+2~jets$
is the same.
The branching ratio $(2/9)\times(6/9)=4/27$ is included.

\begin{center}
\begin{tabular}{l | l l l l l l}\hline\hline
\multicolumn{7}{l}{\sl Electroweak cross-sections in pb;
 left-polarized proton.}\\ \hline
\multicolumn {1}{l|} { Value} &
\multicolumn {6} {c} {Varied quantity}\\ \hline
 & $\Delta\kappa_\gamma$ & $\Delta\kappa_Z$ &
 $\Delta\kappa_{\gamma}=\Delta\kappa_Z$ & $\lambda_\gamma$ &
 $\lambda_Z$ & $\lambda_{\gamma}=\lambda_Z$ \\ \hline
    1.00 &
       1.08$\pm$.02  &   1.88$\pm$.04  &   2.68$\pm$.08  &
       1.26$\pm$.04  &   2.60$\pm$.04  &   3.84$\pm$.08  \\
    0.75 &
       1.02$\pm$.02  &   1.48$\pm$.04  &   1.92$\pm$.04  &
       1.10$\pm$.02  &   1.84$\pm$.04  &   2.60$\pm$.08  \\
    0.50 &
       0.98$\pm$.02  &   1.16$\pm$.02  &   1.34$\pm$.04  &
       1.04$\pm$.02  &   1.38$\pm$.04  &   1.72$\pm$.08  \\
    0.25 &
       0.98$\pm$.02  &   1.00$\pm$.02  &   1.02$\pm$.02  &
       0.98$\pm$.02  &   1.06$\pm$.02  &   1.16$\pm$.02  \\
    0.00 &
       0.96$\pm$.02  &   0.96$\pm$.02  &   0.96$\pm$.02  &
       0.94$\pm$.02  &   0.94$\pm$.02  &   0.98$\pm$.02  \\
   -0.25 &
       0.98$\pm$.02  &   1.08$\pm$.02  &   1.16$\pm$.04  &
       1.00$\pm$.02  &   1.06$\pm$.02  &   1.16$\pm$.04  \\
   -0.50 &
       1.04$\pm$.02  &   1.32$\pm$.02  &   1.58$\pm$.04  &
       1.04$\pm$.02  &   1.40$\pm$.04  &   1.72$\pm$.04  \\
   -0.75 &
       1.14$\pm$.02  &   1.72$\pm$.04  &   2.28$\pm$.04  &
       1.14$\pm$.02  &   1.86$\pm$.04  &   2.60$\pm$.08  \\
   -1.00 &
       1.24$\pm$.04  &   2.24$\pm$.04  &   3.20$\pm$.08  &
       1.26$\pm$.04  &   2.56$\pm$.04  &   3.88$\pm$.08  \\
\hline\hline
\end{tabular}
\end{center}
\newpage
\noindent
Table 3.3.
Purely electroweak cross-sections, in pb, for
$p\bar{p}\rightarrow W^{+}+2~jets$,
with $W^{+}\rightarrow\bar{\ell}\nu$
where $\ell=e$ or $\mu$;
the proton is unpolarized.
The cross-section for $p\bar{p}\rightarrow W^{-}+2~jets$
is the same.
The branching ratio $(2/9)\times(6/9)=4/27$ is included.
These values were calculated independently using the
Morfin-Tung ppdf's;
the cross-section for unpolarized protons is equal to the
average of cross-section for left and right polarized protons.

\begin{center}
\begin{tabular}{l | l l l l l l}\hline\hline
\multicolumn{7}{l}{\sl Electroweak cross-sections in pb;
 unpolarized proton}\\ \hline
\multicolumn {1}{l|} { Value} &
\multicolumn {6} {c} {Varied quantity}\\ \hline
 & $\Delta\kappa_\gamma$ & $\Delta\kappa_Z$ &
 $\Delta\kappa_{\gamma}=\Delta\kappa_Z$ & $\lambda_\gamma$ &
 $\lambda_Z$ & $\lambda_{\gamma}=\lambda_Z$ \\ \hline
    1.00 &
       0.84$\pm$.02  &   1.50$\pm$.04  &   1.88$\pm$.06  &
       1.00$\pm$.02  &   1.98$\pm$.06  &   2.64$\pm$.06  \\
    0.75 &
       0.78$\pm$.02  &   1.14$\pm$.02  &   1.36$\pm$.02  &
       0.86$\pm$.02  &   1.40$\pm$.04  &   1.80$\pm$.06  \\
    0.50 &
       0.74$\pm$.02  &   0.88$\pm$.02  &   0.96$\pm$.02  &
       0.78$\pm$.02  &   1.04$\pm$.02  &   1.22$\pm$.02  \\
    0.25 &
       0.72$\pm$.02  &   0.74$\pm$.02  &   0.76$\pm$.02  &
       0.72$\pm$.02  &   0.80$\pm$.02  &   0.84$\pm$.02  \\
    0.00 &
       0.72$\pm$.02  &   0.72$\pm$.02  &   0.72$\pm$.02  &
       0.70$\pm$.02  &   0.70$\pm$.02  &   0.70$\pm$.02  \\
   -0.25 &
       0.72$\pm$.02  &   0.80$\pm$.02  &   0.84$\pm$.02  &
       0.74$\pm$.02  &   0.86$\pm$.02  &   0.84$\pm$.02  \\
   -0.50 &
       0.78$\pm$.02  &   0.98$\pm$.02  &   1.14$\pm$.02  &
       0.80$\pm$.02  &   1.04$\pm$.02  &   1.22$\pm$.02  \\
   -0.75 &
       0.88$\pm$.02  &   1.28$\pm$.02  &   1.60$\pm$.02  &
       0.88$\pm$.02  &   1.38$\pm$.04  &   1.80$\pm$.06  \\
   -1.00 &
       0.98$\pm$.02  &   1.70$\pm$.04  &   2.28$\pm$.06  &
       1.02$\pm$.02  &   1.92$\pm$.04  &   2.66$\pm$.06  \\
\hline\hline
\end{tabular}
\end{center}
\newpage
\noindent
Table 3.4.
Electroweak cross-sections, in pb, for $W^{+}W^{-}$ production
with non-Standard couplings, with polarized or unpolarized protons,
and with a large-$\sqrt{\hat{s}}$ cut, $\sqrt{\hat{s}}>340$ GeV.
One $W$ decays leptonically, the other to 2 jets, and the branching
ratio $4/27$ is included in the cross-section.
Two assumptions on non-Standard couplings are listed:
$\Delta\kappa_{\gamma}=\Delta\kappa_{Z}$
with $\lambda_{\gamma}=\lambda_{Z}=0$, and
$\lambda_{\gamma}=\lambda_{Z}$
with $\Delta\kappa_{\gamma}=\Delta\kappa_{Z}=0$.

\begin{center}
\begin{tabular}{l | l l l | l l l}\hline\hline
\multicolumn{7}{l}{\sl Electroweak cross-sections in pb, with
$\sqrt{\hat{s}}>340$ GeV}\\ \hline
Value & \multicolumn{6}{c}{Varied parameter}\\ \hline
      & \multicolumn{3}{c|}{$\Delta\kappa_{\gamma}=\Delta\kappa_{Z}$}
 & \multicolumn{3}{c}{$\lambda_{\gamma}=\lambda_{Z}$}\\ \hline
 & $p(L)$ & $p(R)$ & Unpol & $p(L)$ & $p(R)$ & Unpol\\ \hline
    0.75 &
       0.720$\pm$.016  &   0.208$\pm$.004  &   0.456$\pm$.012  &
       1.196$\pm$.028  &   0.316$\pm$.008  &   0.756$\pm$.012  \\
    0.50 &
       0.380$\pm$.008  &   0.110$\pm$.002  &   0.236$\pm$.004  &
       0.592$\pm$.012  &   0.162$\pm$.004  &   0.368$\pm$.008  \\
    0.25 &
       0.180$\pm$.004  &   0.052$\pm$.002  &   0.112$\pm$.002  &
       0.248$\pm$.008  &   0.070$\pm$.002  &   0.152$\pm$.004  \\
    0.00 &
       0.132$\pm$.004  &   0.038$\pm$.002  &   0.082$\pm$.002  &
       0.132$\pm$.004  &   0.038$\pm$.000  &   0.082$\pm$.002  \\
   -0.25 &
       0.228$\pm$.004  &   0.066$\pm$.002  &   0.142$\pm$.004  &
       0.244$\pm$.008  &   0.070$\pm$.002  &   0.156$\pm$.004  \\
   -0.50 &
       0.484$\pm$.012  &   0.138$\pm$.002  &   0.304$\pm$.008  &
       0.604$\pm$.016  &   0.166$\pm$.004  &   0.376$\pm$.008  \\
   -0.75 &
       0.876$\pm$.016  &   0.244$\pm$.004  &   0.556$\pm$.012  &
       1.160$\pm$.028  &   0.316$\pm$.008  &   0.732$\pm$.016  \\
\hline\hline
\end{tabular}
\end{center}
\newpage
\noindent
Table 3.5.
QCD background cross-sections for
$p\bar{p}\rightarrow W^{\pm}+2~jets$,
for various proton polarizations and kinematic cuts.
The unpolarized cases
$\sigma(p\bar{p}\rightarrow W^{+}~2j)=
\sigma(p\bar{p}\rightarrow W^{-}~2j)$
were calculated separately
using CTEQ2 parton distribution functions.

\begin{center}
\begin{tabular}{l l}\hline\hline
\multicolumn{2}{l}{\sl W+2~jet Background}\\ \hline
process & $\sigma$ (pb) \\ \hline
\multicolumn{2}{l}{\sl without 2 jet mass cut: }\\
$p(R)\bar{p}\rightarrow W^-~2j$ & 50.56$\pm$.58\\
$p(L)\bar{p}\rightarrow W^-~2j$ & 41.36$\pm$.52\\
$p(L)\bar{p}\rightarrow W^+~2j$ & 65.46$\pm$.78\\
$p(R)\bar{p}\rightarrow W^+~2j$ & 26.84$\pm$.28\\
$p\bar{p}\rightarrow W^{+}~2j$& 45.30$\pm$.50\\
 \hline
\multicolumn{2}{l}{{\sl with 2 jet mass cut} (70-90 GeV):}\\
$p(R)\bar{p}\rightarrow W^-~2j$ & 7.324$\pm$.070\\
$p(L)\bar{p}\rightarrow W^-~2j$ & 5.910$\pm$.056\\
$p(L)\bar{p}\rightarrow W^+~2j$ & 9.404$\pm$.092\\
$p(R)\bar{p}\rightarrow W^+~2j$ & 3.874$\pm$.034\\
$p\bar{p}\rightarrow W^{+}~2j$& 6.514$\pm$.060\\
 \hline
\multicolumn{2}{l}{\sl with 2 jet mass cut}\\
\multicolumn{2}{l}{{\sl and} $\sqrt{\hat{s}}>340$ GeV:}\\
$p(R)\bar{p}\rightarrow W^{-}~2j$ & 0.348$\pm$.004\\
$p(L)\bar{p}\rightarrow W^{-}~2j$ & 0.308$\pm$.004\\
$p(L)\bar{p}\rightarrow W^{+}~2j$ & 0.520$\pm$.004\\
$p(R)\bar{p}\rightarrow W^{+}~2j$ & 0.144$\pm$.004\\
$p\bar{p}\rightarrow W^{+}~2j$  & 0.358$\pm$.002\\
 \hline\hline
\end{tabular}
\end{center}
\newpage
\noindent
Table 3.6.
Limits on anomalous couplings that could be set
from $p\bar{p}\rightarrow W^{\pm}+2~jets$ with polarized or
unpolarized protons.
The numbers in parentheses are for 10 fb$^{-1}$ integrated
luminosity, and the other numbers are for 1 fb$^{-1}$
integrated luminosity.

\begin{center}
\begin{tabular}{l l l l} \hline\hline
Parameter & \multicolumn{3}{c} {Limits}  \\ \hline
      & $p(R)$ & $p(L)$ & Unpol\\ \hline
$\Delta\kappa_\gamma$ & $\pm 0.89$  & $\pm 1.10$  & $\pm 1.06$ \\
$\Delta\kappa_Z$      & $\pm 0.56$  & $\pm 0.47$  & $\pm 0.54$ \\
$\Delta\kappa_\gamma{=}\Delta\kappa_Z$
                      & $\pm 0.53$  & $\pm 0.35$  & $\pm 0.44$ \\
                      &($\pm 0.30$) &($\pm 0.20$) &($\pm 0.24$)\\
$\lambda_\gamma$      & $\pm 0.79$  & $\pm 0.77$  & $\pm 0.83$ \\
$\lambda_Z$           & $\pm 0.47$  & $\pm 0.37$  & $\pm 0.44$ \\
$\lambda_\gamma{=}\lambda_Z$
                      & $\pm 0.44$  & $\pm 0.29$  & $\pm 0.36$ \\
                      &($\pm 0.25$) &($\pm 0.16$) &($\pm 0.20$)\\
\hline\hline
\end{tabular}
\end{center}
\newpage
\noindent
Table 3.7.
Limits on anomalous couplings that could be set
from $p\bar{p}\rightarrow W^{\pm}+2~jets$ with polarized or
unpolarized protons,
from events with $\sqrt{\hat{s}}>340$ GeV.
The numbers in parentheses are for 10 fb$^{-1}$
integrated luminosity, and the other numbers are for
1 fb$^{-1}$ integrated luminosity.

\begin{center}
\begin{tabular}{l l l l} \hline\hline
Parameter & \multicolumn{3}{c} {\rm Limits} \\ \hline
    & $p(R)$ & $p(L)$ & Unpol \\ \hline
$\Delta\kappa_\gamma{=}\Delta\kappa_Z$
 &  $\pm 0.34$  &  $\pm 0.23$  &  $\pm 0.28$  \\
 & ($\pm 0.19$) & ($\pm 0.13$) & ($\pm 0.16$) \\
$\lambda_\gamma{=}\lambda_Z$
 &  $\pm 0.28$  &  $\pm 0.18$  &  $\pm 0.23$  \\
 & ($\pm 0.16$) & ($\pm 0.10$) & ($\pm 0.13$) \\
 \hline\hline
\end{tabular}
\end{center}
\newpage
\noindent
Table 3.8.
Effect of parton $Q$ scale on the calculated
cross-section for background processes
$p\bar{p}\rightarrow W^{\pm}+2~jets$.
These are unpolarized cross-sections,
calculated with CTEQ2 parton distribution functions.

\begin{center}
\begin{tabular}{l l}\hline\hline
$Q$ scale & $\sigma$ (pb) \\ \hline
$\sqrt{\hat{s}}$   & $6.420\pm .056$\\
$\sqrt{\hat{s}}/2$ & $8.538\pm .074$\\
$2M_W$             & $7.398\pm .068$\\
\hline\hline
\end{tabular}
\end{center}
\newpage

\section*{Figure Captions}
\noindent
Figure 1.1.  Polarized parton distribution functions.
The curves are $x\Delta{f}(x) ~vs~ x$ for parton types
$u_{val}~,~d_{val}~,~u_{sea}(=d_{sea})~,~g$, which
are the most important partons in our calculations.

\noindent
Figure 2.1.
Feynman diagrams for the process
$u\bar{d}\rightarrow W^{+}\gamma$.

\noindent
Figure 2.2.
Total cross-section (with cuts in Eqs. (12)-(15))
for polarized protons {\it vs} anomalous coupling
$\Delta\kappa_{\gamma}$, assuming $\lambda_{\gamma}=0$.
The unpolarized cross-section was calculated separately
using CTEQ2 parton distribution functions.

\noindent
Figure 2.3.
Total cross-section (with cuts in Eqs. (12)-(15))
for polarized protons {\it vs} anomalous coupling
$\lambda_{\gamma}$, assuming $\Delta\kappa_{\gamma}=0$.
The unpolarized cross-section was calculated separately
using CTEQ2 parton distribution functions.

\noindent
Figure 2.4.
Distribution of photon transverse momentum $p_{T\gamma}$.
The solid line is for $\Delta\kappa_{\gamma}=0$,
and the dashed line is for $\Delta\kappa_{\gamma}=-1$;
in both cases $\lambda_{\gamma}=0$.

\noindent
Figure 3.1.
Complete set of electroweak diagrams for
$u+\bar{u}\rightarrow d+\bar{u}+W^{+}
\left(\rightarrow \bar{\ell}+\nu_{\ell}\right)$.

\noindent
Figure 3.2.
Two-jet invariant mass distribution for the signal process.
The solid line is the result of the complete calculation of
the pure electroweak process
$p+\bar{p}\rightarrow q\bar{q}^{\prime}W^{+}$,
with $W^{+}\rightarrow \bar{\ell}\nu_{\ell}$;
the dotted line is the result of the calculation
of $W^{+}W^{-}$ production,
with $W^{+}\rightarrow \bar{\ell}\nu_{\ell}$ and
$W^{-}\rightarrow 2~jets$, with a cut on the
two-jet invariant mass ($70<M_{2j}<90$ GeV).

\noindent
Figure 3.3.
Two-jet invariant mass distribution for the QCD background
processes $p\bar{p}\rightarrow W^{+}+2~jets$.
(The cross-section for $W^{-}+2~jets$ is the same,
for unpolarized scattering.)

\noindent
Figure 3.4.
$\sqrt{\hat{s}}$ distribution for the signal process.
The solid line is the result of the complete calculation of
the pure electroweak process
$p+\bar{p}\rightarrow q\bar{q}^{\prime}W^{+}$,
with $W^{+}\rightarrow \bar{\ell}\nu_{\ell}$;
the dotted line is the result of the calculation
of $W^{+}W^{-}$ production,
with $W^{+}\rightarrow \bar{\ell}\nu_{\ell}$ and
$W^{-}\rightarrow 2~jets$, with a cut on the
two-jet invariant mass ($70<M_{2j}<90$ GeV).

\noindent
Figure 3.5.
Comparison of $\sqrt{\hat{s}}$ distributions for
signal and background processes.
The solid line is the QCD background.
The dotted line is the electroweak process, with
zero anomalous couplings; the dashed line is the
electroweak process with
$\Delta\kappa_{\gamma}=\Delta\kappa_{Z}=0.5$.

\noindent
Figure 3.6.
Comparison of $\sqrt{\hat{s}}$ distributions for
signal and background processes,
for $\sqrt{\hat{s}}>340$ GeV.
The solid line is the QCD background.
The dotted line is the electroweak process, with
zero anomalous couplings; the dashed line is the
electroweak process with
$\Delta\kappa_{\gamma}=\Delta\kappa_{Z}=0.5$.

\noindent
Figure 3.7.
Electroweak cross-section for $p_{\lambda}\bar{p}\rightarrow W^{+}W^{-}$
as a function of anomalous couplings $\Delta\kappa$,
for polarized $(L,R)$ and unpolarized protons.
For each polarization, three cases are shown, corresponding to
assumptions
($\times$) $\Delta\kappa_{\gamma}\neq 0$ and $\Delta\kappa_{Z}=0$,
($\Diamond$) $\Delta\kappa_{\gamma}=0$ and $\Delta\kappa_{Z}\neq 0$,
and ($\Box$) $\Delta\kappa_{\gamma}=\Delta\kappa_{Z}$.
In all cases $\lambda_{\gamma}=\lambda_{Z}=0$.

\noindent
Figure 3.8.
Electroweak cross-section for $p_{\lambda}\bar{p}\rightarrow W^{+}W^{-}$
as a function of anomalous couplings $\lambda$,
for polarized $(L,R)$ and unpolarized protons.
For each polarization, three cases are shown, corresponding to
assumptions
($\times$) $\lambda_{\gamma}\neq 0$ and $\lambda_{Z}=0$,
($\Diamond$) $\lambda_{\gamma}=0$ and $\lambda_{Z}\neq 0$,
and ($\Box$) $\lambda_{\gamma}=\lambda_{Z}$.
In all cases $\Delta\kappa_{\gamma}=\Delta\kappa_{Z}=0$.

\noindent
Figure 3.9.
Electroweak cross-section for $p_{\lambda}\bar{p}\rightarrow W^{+}W^{-}$
as a function of anomalous couplings,
with $\sqrt{\hat{s}}>340$ GeV.
For each polarization, two cases are shown, corresponding to
assumptions
($\times$) $\Delta\kappa_{\gamma}=\Delta\kappa_{Z}$ with
$\lambda_{\gamma}=\lambda_{Z}=0$,
and ($\Diamond$) $\lambda_{\gamma}=\lambda_{Z}$ with
$\Delta\kappa_{\gamma}=\Delta\kappa_{Z}=0$.

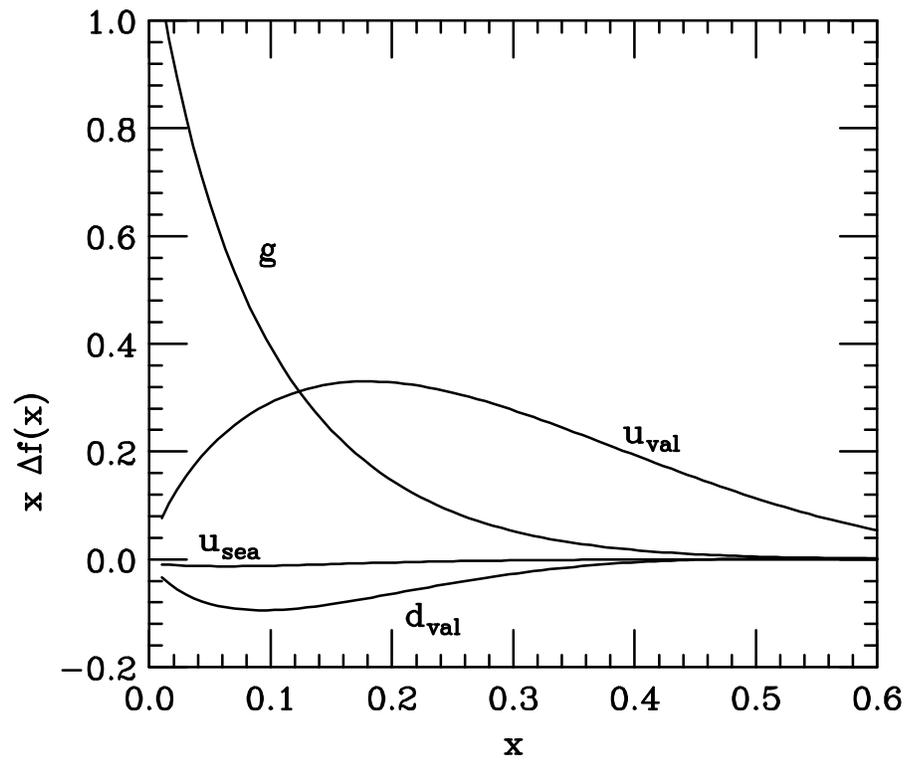
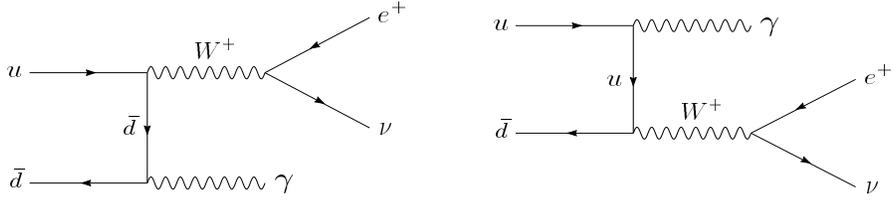
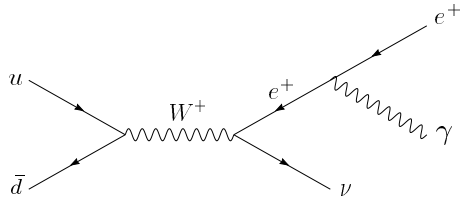
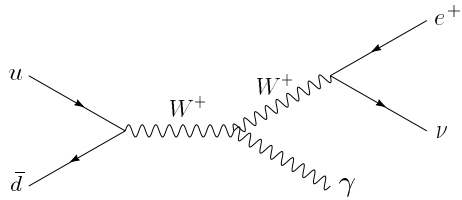
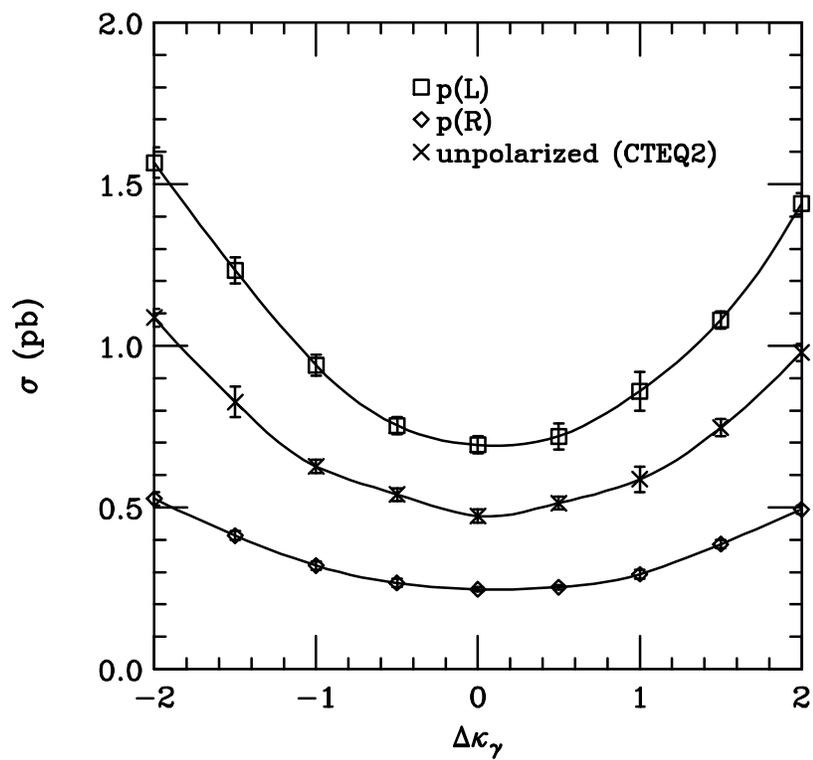
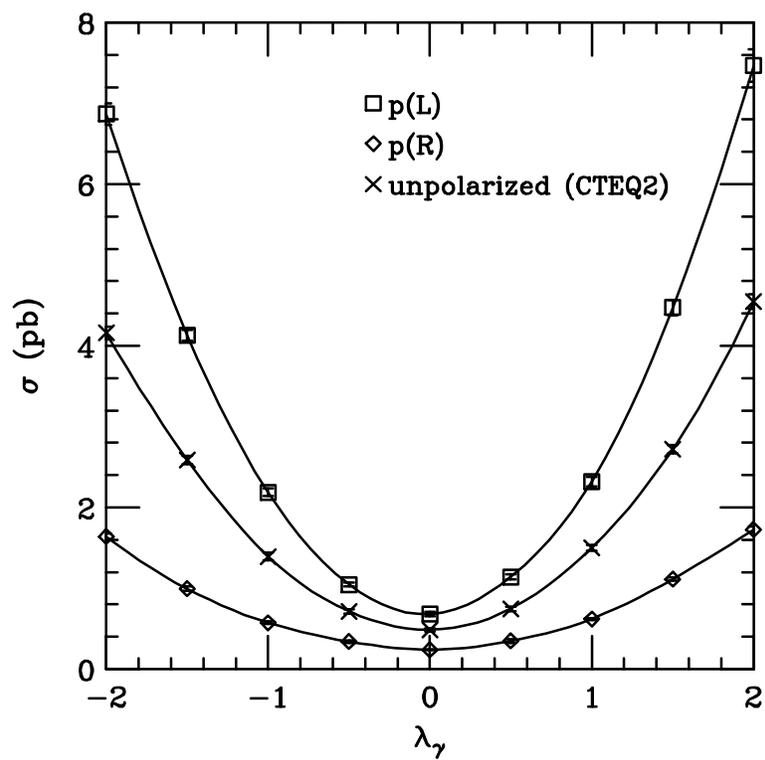
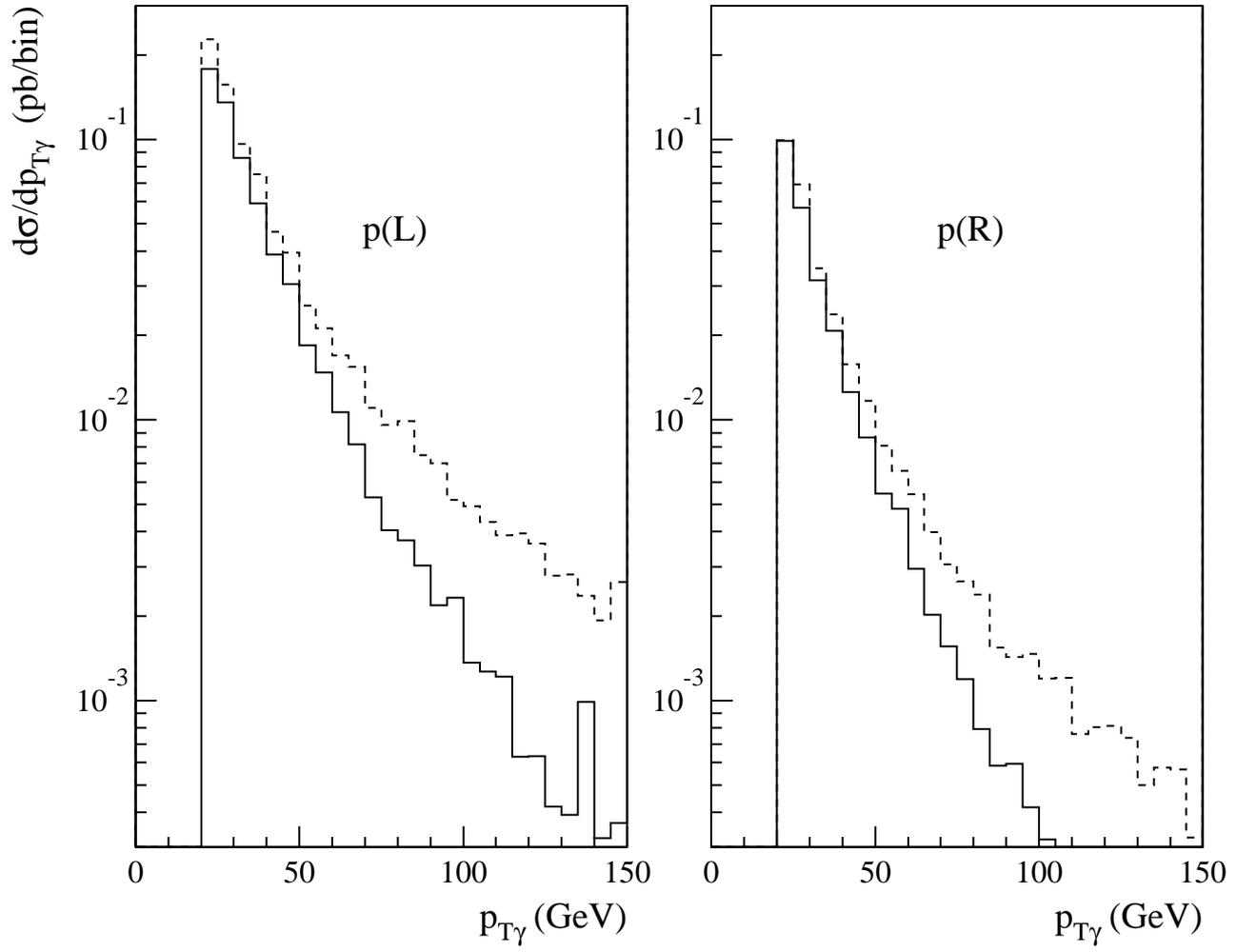
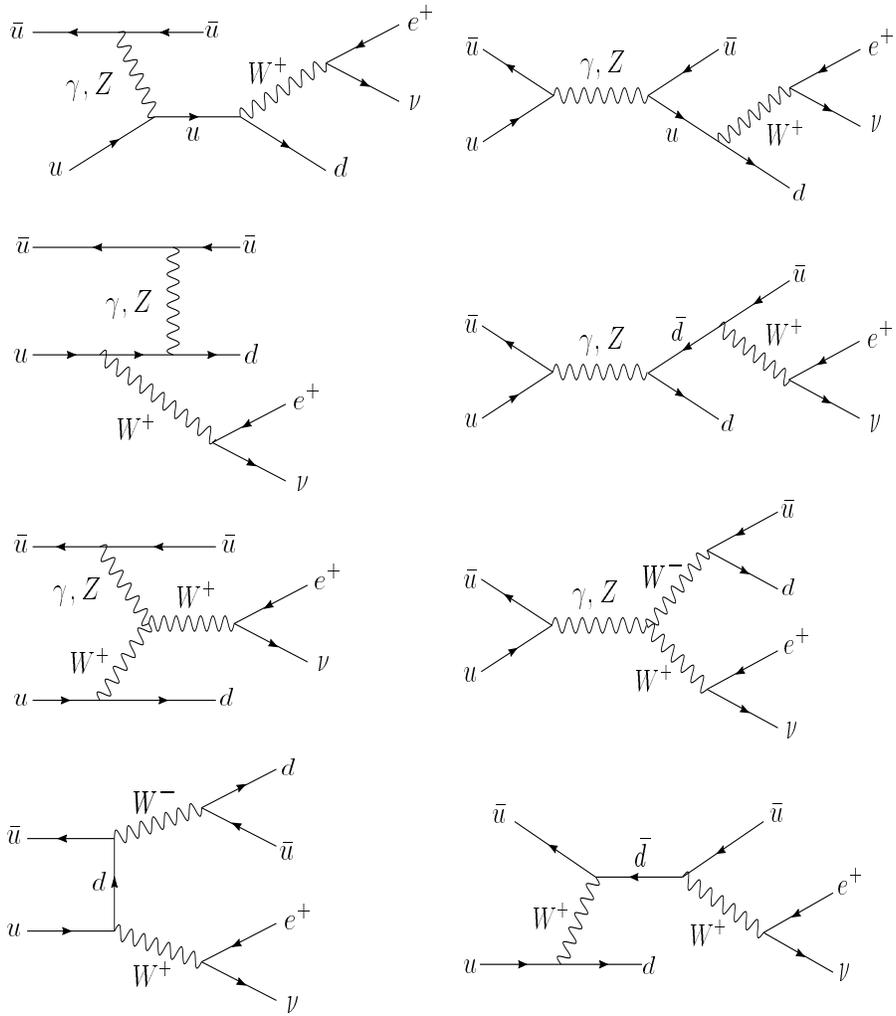
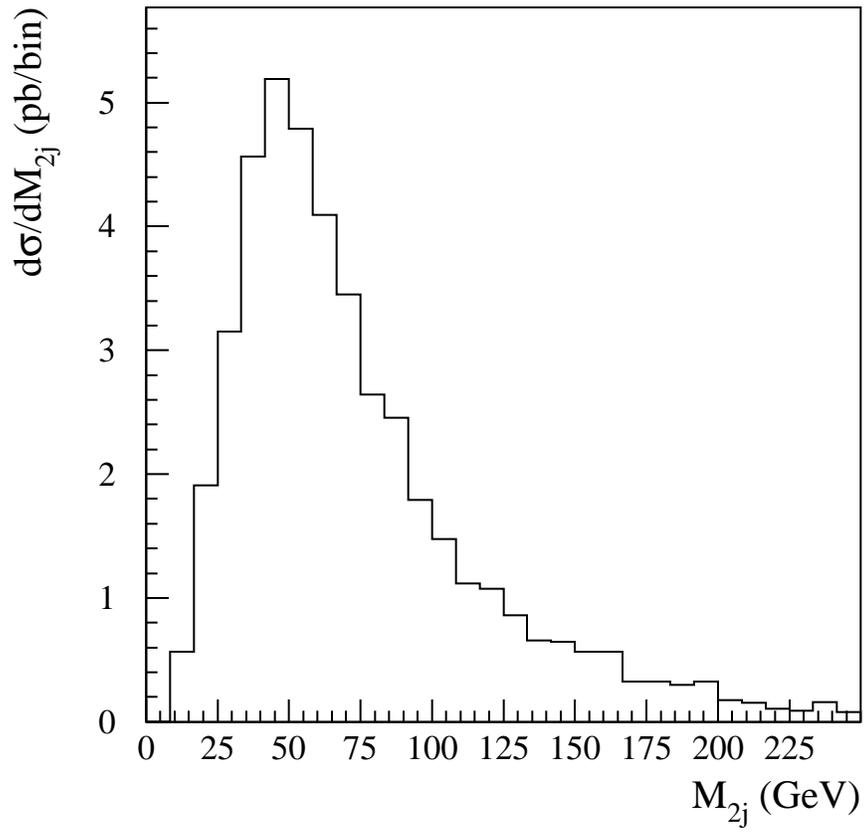
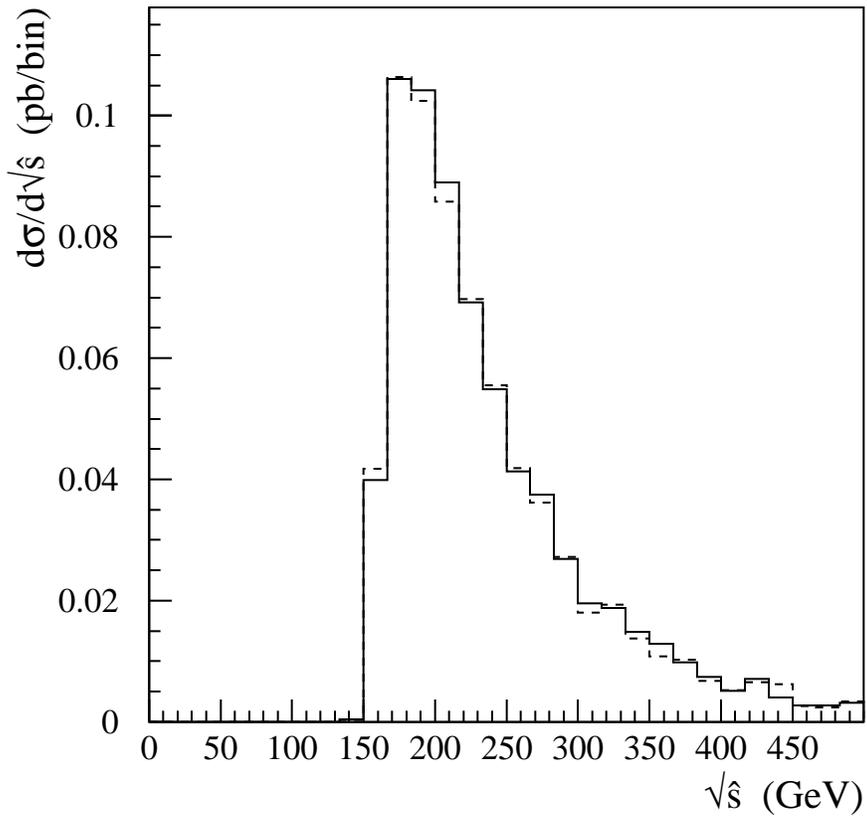
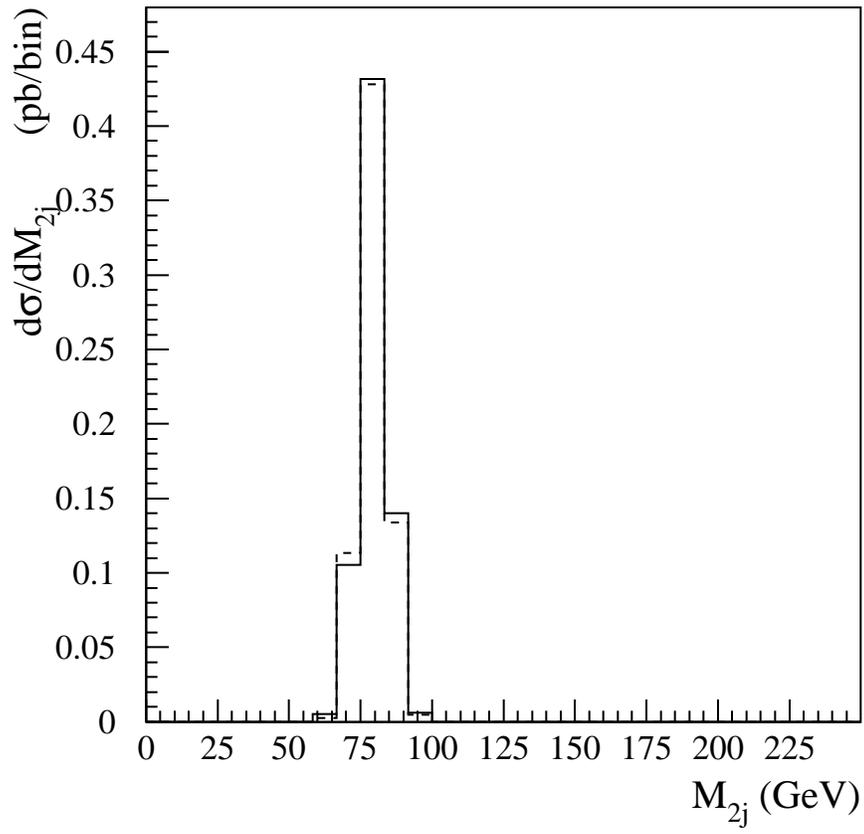
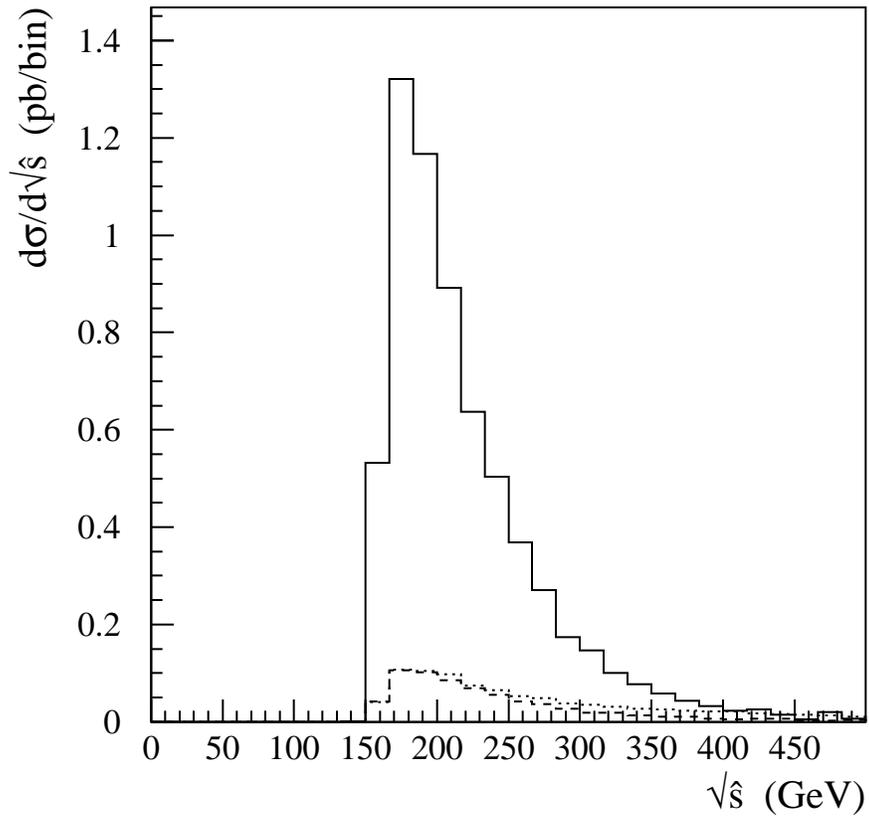
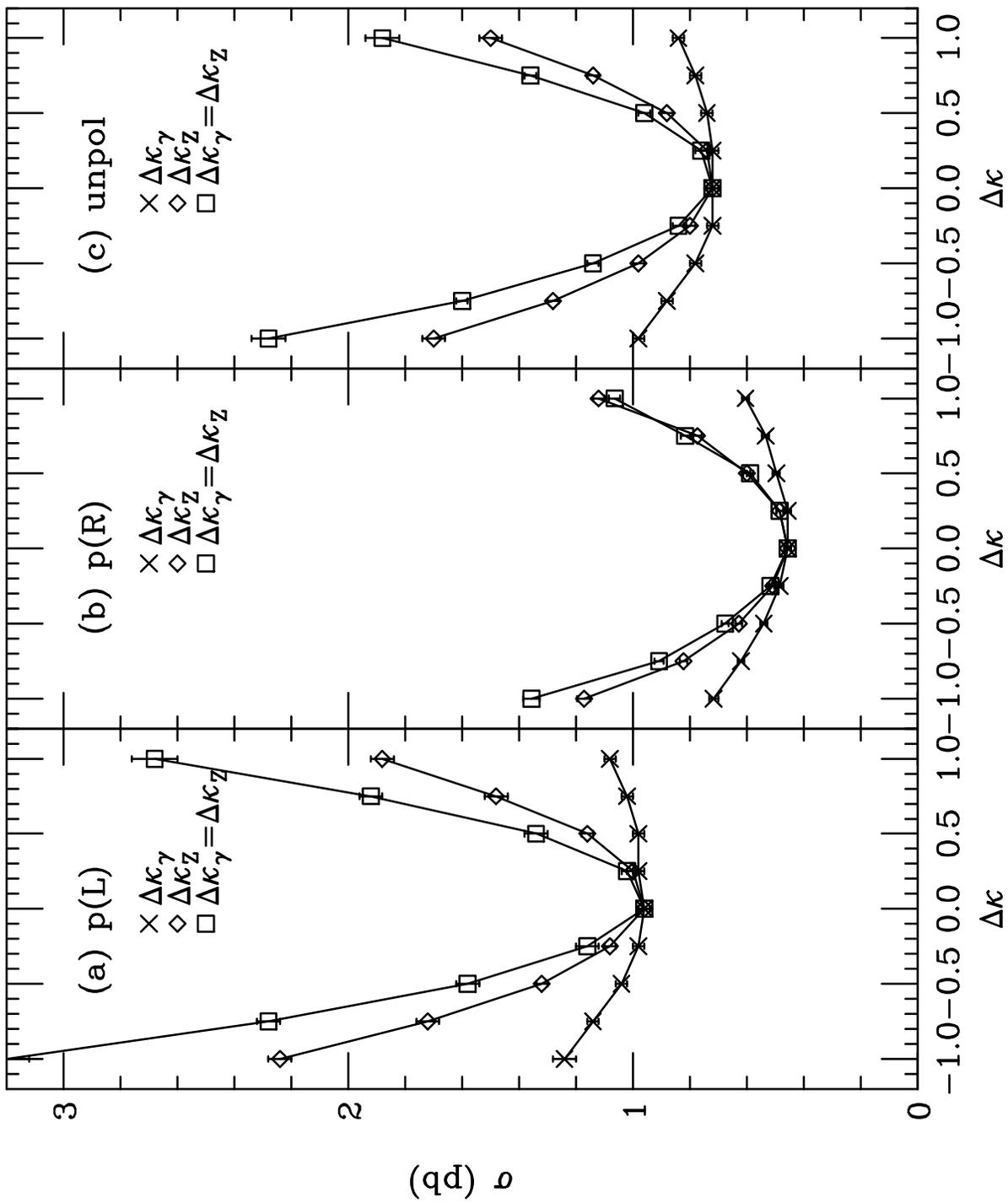
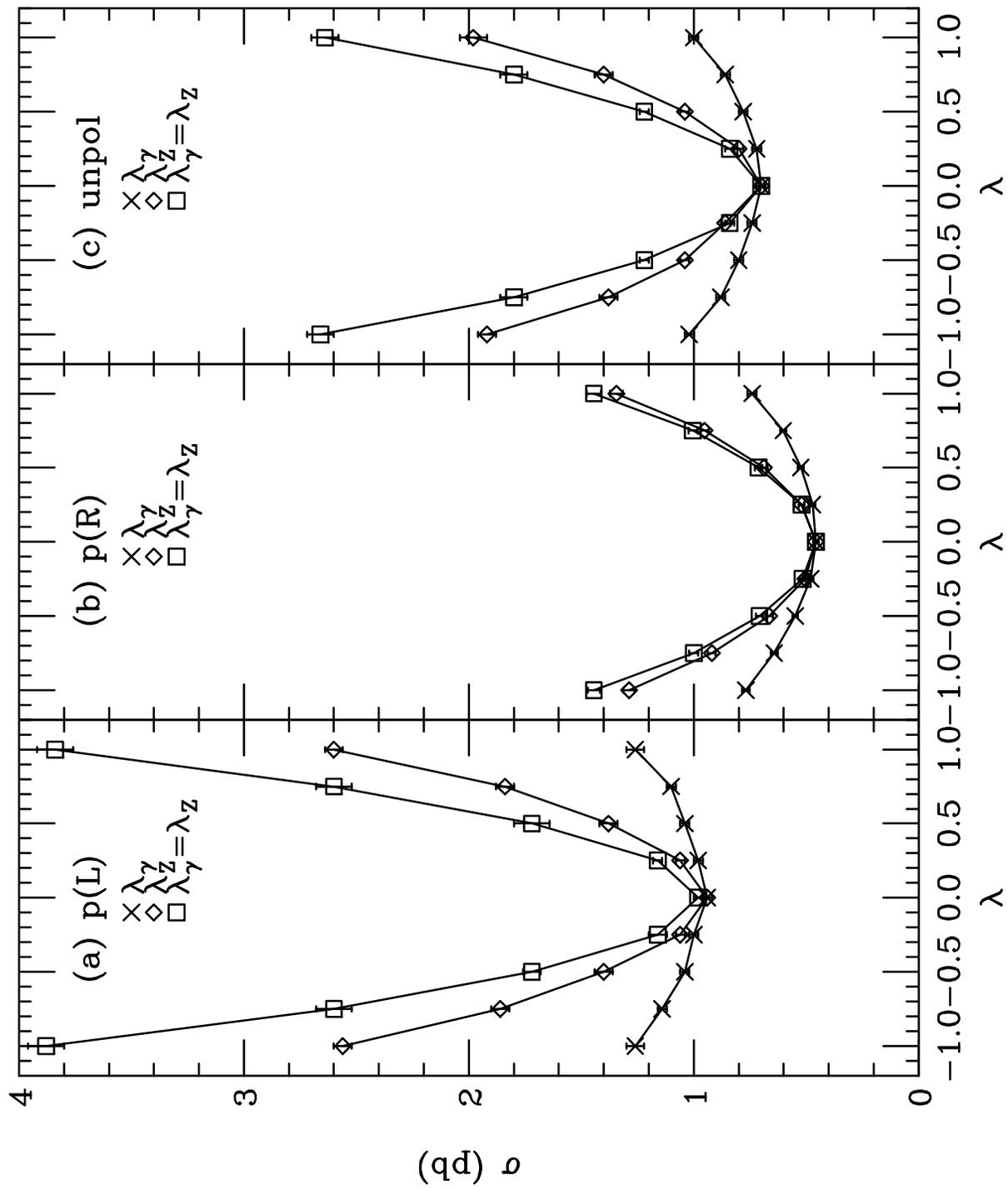
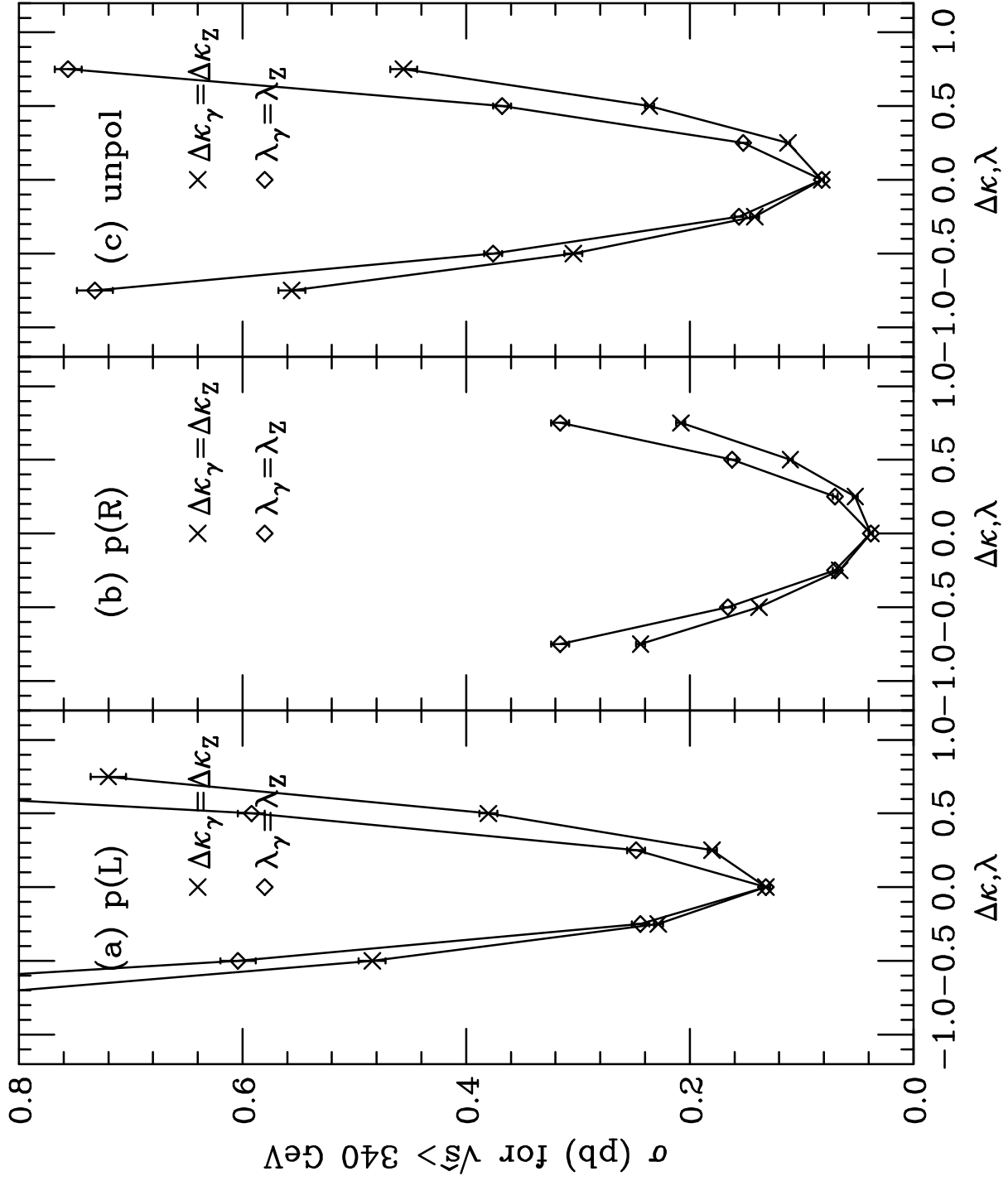
\end{document}